\documentstyle[epsfig]{mn}

\newif\ifAMStwofonts

\ifoldfss
  \newcommand{\rmn}[1] {{\rm #1}}

  \ifCUPmtlplainloaded \else
    \NewTextAlphabet{textbfit} {cmbxti10} {}
    \NewTextAlphabet{textbfss} {cmssbx10} {}
    \NewMathAlphabet{mathbfit} {cmbxti10} {} 
    \NewMathAlphabet{mathbfss} {cmssbx10} {} 
  \fi
  \ifAMStwofonts
    \ifCUPmtlplainloaded \else
      \NewSymbolFont{upmath} {eurm10}
      \NewSymbolFont{AMSa} {msam10}
      \NewMathSymbol{\upi}     {0}{upmath}{19}
      \NewMathSymbol{\umu}     {0}{upmath}{16}
      \NewMathSymbol{\upartial}{0}{upmath}{40}
      \NewMathSymbol{\leqslant}{3}{AMSa}{36}
      \NewMathSymbol{\geqslant}{3}{AMSa}{3E}

      \let\leq=\leqslant 
      \let\geq=\geqslant 
    \fi
  \fi
\fi 

\ifnfssone
  \newmathalphabet{\mathit}
  \addtoversion{normal}{\mathit}{cmr}{m}{it}
  \addtoversion{bold}{\mathit}{cmr}{bx}{it}
  \newcommand{\rmn}[1] {\mathrm{#1}}

  \newmathalphabet{\mathbfit} 
  \addtoversion{normal}{\mathbfit}{cmr}{bx}{it}
  \addtoversion{bold}{\mathbfit}{cmr}{bx}{it}
  \newmathalphabet{\mathbfss} 
  \addtoversion{normal}{\mathbfss}{cmss}{bx}{n}
  \addtoversion{bold}{\mathbfss}{cmss}{bx}{n}
  \ifAMStwofonts
    \ifCUPmtlplainloaded \else
      \UseAMStwoboldmath
      \makeatletter
      \new@mathgroup\upmath@group
      \define@mathgroup\mv@normal\upmath@group{eur}{m}{n}
      \define@mathgroup\mv@bold\upmath@group{eur}{b}{n}
      \edef\UPM{\hexnumber\upmath@group}
      \new@mathgroup\amsa@group
      \define@mathgroup\mv@normal\amsa@group{msa}{m}{n}
      \define@mathgroup\mv@bold\amsa@group{msa}{m}{n}
      \edef\AMSa{\hexnumber\amsa@group}
      \makeatother
      \mathchardef\upi="0\UPM19
      \mathchardef\umu="0\UPM16
      \mathchardef\upartial="0\UPM40
      \mathchardef\leqslant="3\AMSa36
      \mathchardef\geqslant="3\AMSa3E

      \let\leq=\leqslant 
      \let\geq=\geqslant 
    \fi
  \fi
\fi 

\ifnfsstwo
  \newcommand{\rmn}[1] {\mathrm{#1}}

  \DeclareMathAlphabet{\mathbfit}{OT1}{cmr}{bx}{it}
  \SetMathAlphabet\mathbfit{bold}{OT1}{cmr}{bx}{it}
  \DeclareMathAlphabet{\mathbfss}{OT1}{cmss}{bx}{n}
  \SetMathAlphabet\mathbfss{bold}{OT1}{cmss}{bx}{n}
  \ifAMStwofonts
    \ifCUPmtlplainloaded \else
      \DeclareSymbolFont{UPM}{U}{eur}{m}{n}
      \SetSymbolFont{UPM}{bold}{U}{eur}{b}{n}
      \DeclareSymbolFont{AMSa}{U}{msa}{m}{n}
      \DeclareMathSymbol{\upi}{0}{UPM}{"19}
      \DeclareMathSymbol{\umu}{0}{UPM}{"16}
      \DeclareMathSymbol{\upartial}{0}{UPM}{"40}
      \DeclareMathSymbol{\leqslant}{3}{AMSa}{"36}
      \DeclareMathSymbol{\geqslant}{3}{AMSa}{"3E}

      \let\leq=\leqslant 
      \let\geq=\geqslant 
    \fi
  \fi
\fi 

\ifCUPmtlplainloaded \else
  \ifAMStwofonts \else 
    \def\upi{\pi}
    \def\umu{\mu}
    \def\upartial{\partial}
  \fi
\fi

\title{J-type Carbon Stars in the Large Magellanic Cloud} 
\author[D.H. Morgan et al.]
       {D.H.~Morgan,$^1$ R.D.~Cannon,$^2$ D.~Hatzidimitriou$^3$
        B.F.W.~Croke$^4$ \\
       $^1$ Institute for Astronomy, University of Edinburgh, Royal 
            Observatory, Blackford Hill, Edinburgh  EH9 3HJ, UK \\
       $^2$ Anglo-Australian Observatory, PO Box 296, Epping, NSW 2121, 
            Australia  \\
       $^3$ Department of Physics, University of Crete, Heraklion, Greece \\
       $^4$ Integrated Catchment Assessment and Management Centre,
            Centre for Resource and Environment Studies, \\
            Australian National University, Canberra, ACT 0200, Australia}
\date{Accepted 
      Received 
      in original form }

\pagerange{\pageref{firstpage}--\pageref{lastpage}}
\pubyear{2000}

\begin{document}

\maketitle

\label{firstpage}

\begin{abstract}
A sample of 1497 carbon stars in the Large Magellanic Cloud has been 
observed in the red part of the spectrum with the 2dF facility on 
the AAT.  Of these, 156 have been identified as J-type 
(i.e. $^{13}{\rmn C}$-rich) carbon stars using a 
technique which provides a clear distinction between J~stars and the
normal N-type carbon stars that comprise the bulk of the sample,
and yields few borderline cases.   A simple 2-D classification of the 
spectra, based on their spectral slopes in different wavelength 
regions, has been constructed and found to be 
related to the more conventional c- and j-indices, modified to suit the 
spectral regions observed.  Most of the J~stars form a photometric 
sequence in the $K - (J-K)$ colour magnitude diagram, parallel to 
and 0.6\,mag fainter than the N~star sequence.  A subset of the 
J~stars (about 13 per cent) are brighter than this J-star sequence; 
most of these are spectroscopically different from the other J~stars.   
The bright J~stars have stronger CN bands than the other J~stars and 
are found strongly concentrated in the central regions of the LMC.
Most of the rather few stars in common with Hartwick and Cowley's sample of
suspected CH~stars are J~stars.  Overall, the proportion of carbon stars 
identified as J~stars is somewhat 
lower than has been found in the Galaxy.  The Na\,D lines are weaker in
the LMC J~stars than in either the Galactic J~stars or the LMC N~stars, 
and do not seem to depend on temperature.  

\end{abstract}


\section{Introduction}

The green-red optical spectra of carbon stars are dominated by the Swan 
bands of $^{12}{\rmn C}^{12}{\rmn C}$ and bands from the red 
A\,$^2\Pi$--X\,$^2\Sigma$ $^{12}{\rmn C}^{14}{\rmn N}$ system.
Also present in some carbon stars are isotopic bands from the 
$^{13}{\rmn C}^{12}{\rmn C}$ and $^{13}{\rmn C}^{14}{\rmn N}$ molecules:
stars showing strong isotopic bands are known as J-type carbon stars.  
In most carbon stars the molecular bands are so strong that they obscure most 
atomic lines in all but very high-resolution spectra.  Despite this,
carbon stars are recognised in four principal spectroscopic groups
\cite{keenan93} -- R, N, CH and J, plus several rarer types such as
hydrogen-deficient carbon stars (H-d).  The basic characteristics of the 
R~stars are relatively strong fluxes in the blue-violet and moderately strong 
isotopic bands, whereas the N~stars have heavy diffuse absorption 
in the blue, weak or absent isotopic bands, and show enhancements in
the strengths of the lines of light s-process elements such as Ba.  
The CH~stars are characterised by strong bands from the CH~molecule 
and the J~stars by strong $^{13}{\rmn C}$\, isotopic bands.  Abundance 
analyses show that J~stars have $^{12}{\rmn C}$\,:$^{13}{\rmn C}$ 
ratios as low as 4, whereas the common cool N~stars have 
$^{12}{\rmn C}$\,:$^{13}{\rmn C}$ ratios between 40 and the solar value 
of 80 \cite{lambert86}.  Like the R~stars, the J~stars do not show 
enhanced lines from the s-process elements.  Some J~stars exhibit 
infrared emission from oxygen-rich dust shells \cite{evans90}.

The types of Galactic carbon star are also associated with different 
stellar populations.  According to Barnbaum, Stone \& Keenan 
\shortcite{barnbaum96}), Galactic N~stars are bright 
giants with $M_V\simeq-2.2$ belonging to the thin disk, the R~stars 
are normal giants with $M_V\simeq0$ belonging to the thicker disk, 
indicating that they probably arise from lower mass, older 
progenitors, while most CH~stars are bright with $M_V\simeq-1.8$ and 
belong to the Galactic halo, i.e., old Population II.  In contrast, 
relatively little is known of the J~stars.  
Abia \& Isern \shortcite{abia00} have
concluded that the evolutionary status of the J~stars
is still uncertain in that standard evolutionary AGB models are unable to
explain properly the observed chemical properties of the J~stars; they have
also discussed various mechanisms for providing additional mixing within 
the stars.   They mention the possible existence of two groups of 
Galactic J~stars with different luminosities arising from stars of different 
initial mass and perhaps following different evolutionary mechanisms.
 
The stellar content of the Large Magellanic Cloud (LMC) includes a 
large population of about 8000 observable carbon stars \cite{kontizas01b} 
which, being at a known distance and suffering from very little interstellar 
reddening, provides an excellent source for studying the photometric 
and spectroscopic properties of carbon stars.  Carbon stars are much 
more common in the LMC than in the Galaxy, as measured by the ratio of 
C to late-M stars which is 2.2 in the LMC and $\leq$0.001 in the 
Galactic Bulge \cite{blanco83}; this is believed to be a consequence 
of the overall lower metallicity in the LMC, although a difference in 
star formation history may also be critical.  Until now, most of 
the LMC J~stars known were those identified as
such by Richer, Olander \& Westerlund \shortcite{richer79} from an early 
catalogue of carbon stars \cite{westerlund78} which naturally selected 
the brightest carbon stars.   Some fainter J~stars had been found and 
seen to lie photometrically near the lower envelope of the N~star 
distribution (Richer \shortcite{richer81}, Bessell, Wood \& Lloyd Evans 
\shortcite{bessell83} and Westerlund et al. \shortcite{westerlund91}), 
but in all these cases
the number of faint J~stars was small.  Hence, it was thought that the 
J~stars were brighter than the N~stars, but with a few faint exceptions.  
It is important to study the J-star population in the LMC in greater 
numbers and down to fainter magnitudes to determine how these stars 
are related to the other types of carbon star both photometrically and 
kinematically.

In order to address these and other issues, an extensive programme of 
spectroscopy of LMC carbon stars has been started, using 
the 2dF multi-object spectroscopy facility on the Anglo-Australian 
Telescope (AAT) \cite{cannon99}.  The new dataset includes more than 
150 J~stars.  This paper describes these spectra.  It is organised as 
follows: Section~2 describes the observations and Section~3 describes 
how the sample of J~stars was selected.  The J~stars by no means form 
a homogeneous set of spectra.  Consequently, they are divided into 
groups according to their overall spectral appearance.  This is done 
in Section~4.  Photometry in $RIJHK$ is available for most of the stars 
and is described in Section~5.  Section~6 is a more detailed spectral 
study of the stars, and some individual stars are described in Section 7.
The overall results are discussed in Section~8.

\section{Observations}

Most of the carbon stars observed with 2dF were selected from the newly 
completed catalogue of 7760 carbon stars by Kontizas et al. 
\shortcite{kontizas01b} -- KDMK01 (see also Dapergolas et al. 
\shortcite{dapergolas96}).  This catalogue was constructed by identifying 
the strong (1,\,0) and (0,\,0) Swan bands of carbon stars seen in 
low-dispersion spectra on an objective-prism survey taken in the 
blue-green with the 1.2\,m UK Schmidt Telescope (UKST).   

The 2dF instrument on the AAT enables up to 400 spectra to be recorded
simultaneously, in a pair of identical fibre-fed spectrographs fitted 
with TEK~1024 CCDs \cite{lewis02} -- see http://www.aao.gov.au/2df/.   
Most of the spectra for 
this programme were obtained during January and November 1998.  
These spectra were taken in the red, centred on the $\Delta\nu$\,=\,+2 
Swan bands near 6200\AA.  The first set of observations covered 
four 2-degree diameter fields in the wavelength range 5675--6785\AA, 
and the second set gave five fields in the wavelength range 5585--6700\AA.  
Thus, the strong (0,\,1) Swan band at 5635\AA \/ was included in the 
second set of observations but not in the first set, whereas the
latter included the Li\,{\sc i}\,$\lambda$6708 line which is the subject 
of a separate paper \cite{hatzidimitriou03}.  One field was 
common to both sets of observations to provide checks on the system.  
Most of the stars observed fall in the brightness range $R\sim14-16$.  
Altogether, $\sim$1600 carbon star spectra were obtained on eight field 
centres arranged approximately on a 10$^\circ$ long north-south strip 
across the central parts of the LMC.   

The 1200R gratings were used, yielding spectra with 1.1\AA \/ per pixel
or an effective resolution of $\sim2.5$\AA.  Several exposures were taken 
for each field, typically 3\,$\times$\,900\,s or 2\,$\times$\,1200\,s,
together with offset sky, arc and flat field exposures.
The data were reduced with the AAO's {\sc 2dfdr} data reduction package 
\cite{bailey01}
and the {\sc iraf} package, and yielded spectra with signal-to-noise 
ratio (S/N) usually around 30 per pixel.  Full details of the observations 
are given by Cannon et al. (in peparation).

One central LMC field was re-observed in January 2002, using the 1200B
gratings to give blue spectra in the wavelength range 4190--5290\AA \/ 
at 1.1\AA \/ per pixel.  This was done as part of the 
ATAC Service Observing Scheme.

Radial velocities of all the stars were obtained by using a 
cross-correlation technique in {\sc figaro}.  The high S/N
usually attained for most stars, typically around 30 per pixel, and the 
large number of spectral features resulted in internal accuracies of 
a few km\,s$^{-1}$, although external errors are larger.

\begin{figure}
\vspace{11.5cm}
\includegraphics{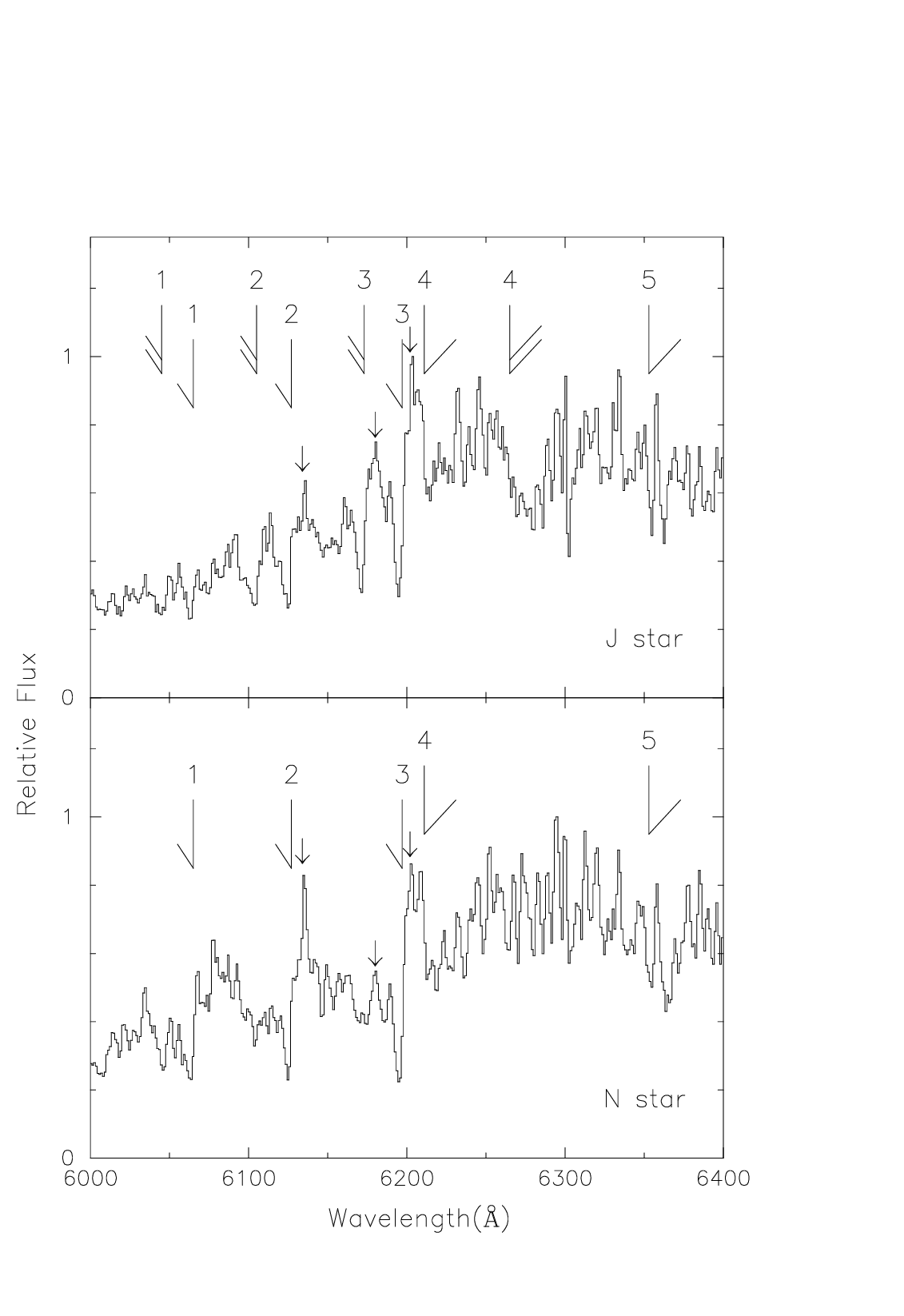}
\caption{2dF spectra of a J~star and an N~star in the 6000--6400\AA \/ region,
scaled to a maximum of unity (see text for details).
The main molecular bands are marked and numbered as follows:
(1) (2,\,4) C$_2$, (2) (1,\,3) C$_2$,  (3) (0,\,2)  C$_2$, 
(4) (4,\,0) CN and (5) (5,\,1) CN.  The isotopic bands are identified
by the double bar line.  The arrows mark three key
continuum points used in Section~4.}
\label{plex}
\end{figure}

\begin{figure}
\vspace{10.5cm}
\includegraphics{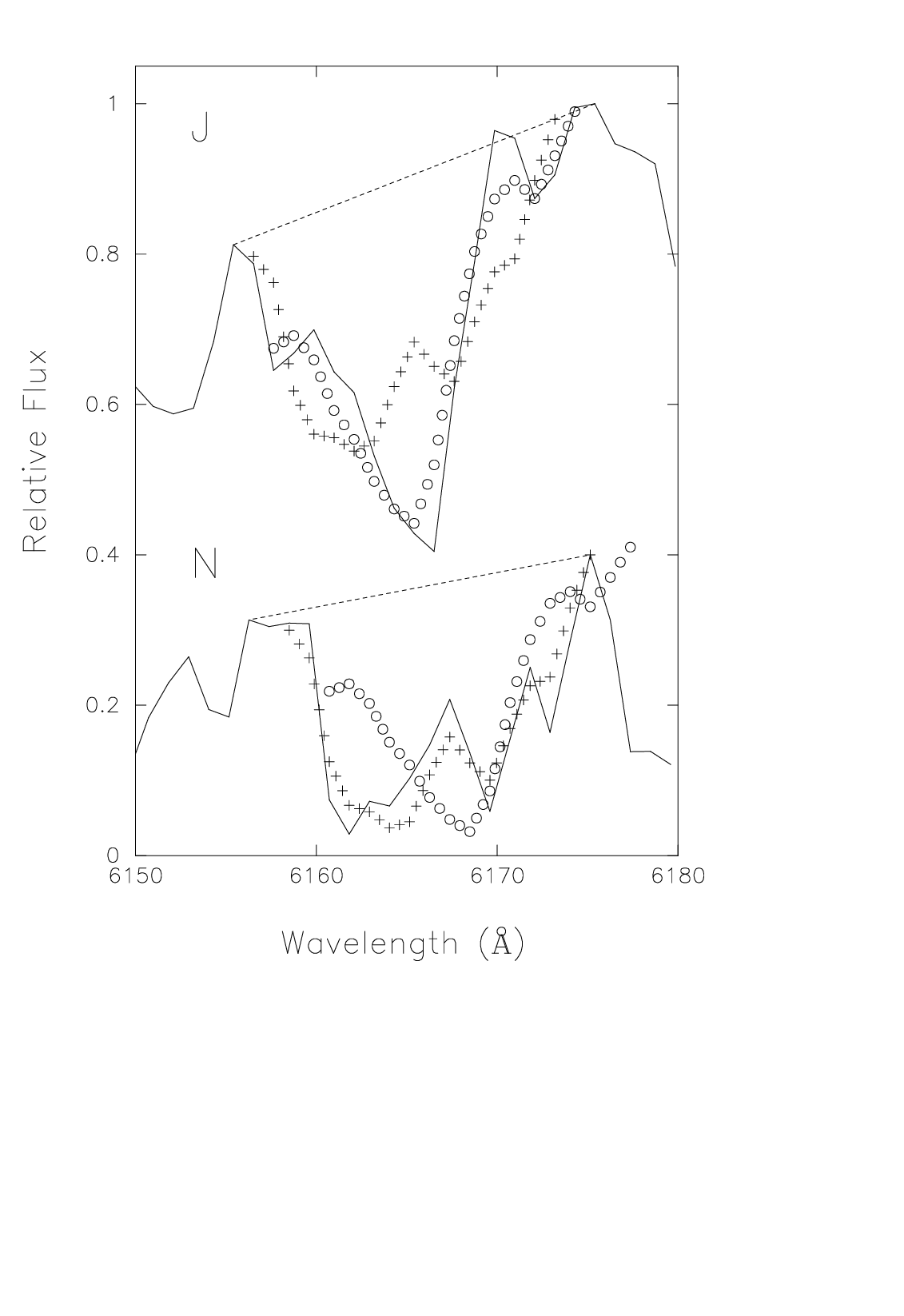}
\caption{Fitting the spectra around 6165\AA.  The upper spectrum 
(continuous line) is of a typical J~star and the lower one, which is 
offset downwards by 0.6, is of a typical N~star.  The lines marked by 
the plus signs are the best fits for the N-star template and by the 
circles are the best fits for the J-star template (see text for details).}
\label{sp_sift}
\end{figure}

\begin{figure}
\vspace{12.0cm}
\includegraphics{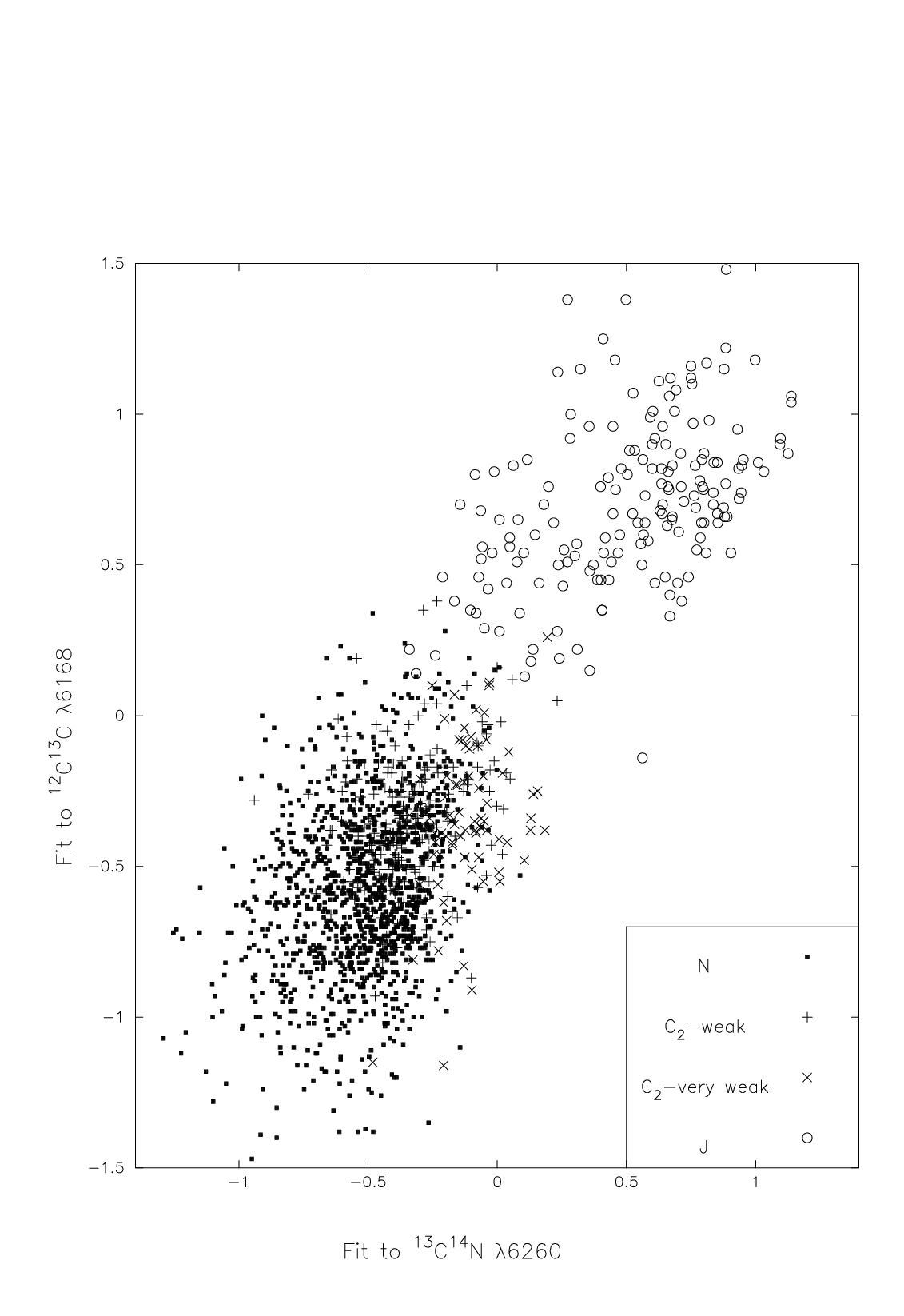}
\caption{Selection of the J~stars.  The axes are measures of the difference 
between the individual spectra and templates of N and J~stars (see text).  
The plotted values are positive when the spectrum matches the J~star better
than the N~star and negative for the reverse.  Otherwise the scale
is arbitary.  The abscissa relates to the (4,\,0) CN bands and the
ordinate to the 6159--6175\AA \/ spectral region which includes the
$^{13}{\rmn C}^{12}{\rmn C}~\lambda6168$ band (see Fig.\ref{sp_sift}).  
The symbols are as shown in the inset.}
\label{jfind_b}
\end{figure}

\section{J~star identification}

Fig.~\ref{plex} shows sections of representative spectra of J and N ~stars 
in the 6000-6400\AA \/ waveband and is a good illustration of the quality 
of the 2dF spectra.  The plots are the reduced spectra after wavelength 
calibration and sky subtraction but with no intensity or flux calibration 
(for details see Cannon et al. in preparation), and are scaled to their 
maxima within the plotted waveband.  The principal molecular bands are
marked on the figure.  It 
was straightforward to make a visual inspection of the spectra and thereby
note as J~stars those stars with an obvious $^{13}{\rmn C}^{12}{\rmn C}$ 
band at 6168\AA.  This inspection also allowed a simple three-fold 
categorisation of the remaining stars according to whether the strength 
of the C$_2$ bands was strong, weak or very weak. The first of these 
groups typifies the common, cool N~stars.

One of the difficulties in identifying 
$^{13}{\rmn C}^{12}{\rmn C}~\lambda6168$ when 
it is weak is the proximity of Ca\,{\sc i}\,$\lambda$6162 and various CN and 
$^{12}{\rmn C}^{12}{\rmn C}$ bands.  In order to make a systematic 
identification of J~stars, the spectra were first adjusted to zero 
radial velocity using the measured radial velocities.  In most 
carbon stars, the spectrum between 6159\AA \/ and 6175\AA \/ shows 
a broad absorption feature with a minimum at 6162--6164\AA \/ and a 
secondary minimum at 6158\AA; in contrast, the J~stars show a single 
minimum at 6168\AA.  (In stars where C$_2$ is weak, the individual 
components are well separated with the Ca\,{\sc i}\,$\lambda$6162 line 
being the 
stronger.)  This feature was then compared with two templates, one being 
a mean of stars already deemed to be normal carbon stars (N~stars) and 
the other a mean of stars deemed to be J~stars.  These templates were 
constructed from stars in one field observed with CCD1 and included
115 N~stars and 20 J~stars.  The spectra and templates were all normalized to 
local continua defined by the fluxes near 6156\AA \/ and 6175\AA.  
By allowing the templates to be shifted in wavelength to allow for 
pixellation differences and then scaled in strength, the best fit of 
each template to each star was taken to be the minimum rms of the 
pixel-to-pixel difference between the stellar spectrum and the 
template ($\sigma_{1,N}$ or $\sigma_{1,J}$).  Fig.~\ref{sp_sift} 
shows an example of these fits for a typical J~star (KDM\,2348) and a 
typical N~star (KDM\,2109).  The normalising continua are indicated
by dashed lines.

J~stars also show the (4,\,0) $^{13}{\rmn C}^{14}{\rmn N}$ band at 6260\AA \/ 
as well as the (4,\,0) $^{12}{\rmn C}^{14}{\rmn N}$ band at 6206\AA.
A procedure similar to the above was applied to this (much larger)
waveband by fitting templates of the two types in the 6205-6290\AA \/ 
waveband.  This was done twice for each template, first by maintaining 
the shape of the template throughout the 
band and secondly by allowing the scaling factor to be different in the
two halves 6205-6250\AA \/ and 6255-6290\AA.  The splitting of the band was
necessary to account for J~stars with different relative strengths of these
two isotopic bands.  There is, of course, no benefit to be gained from 
splitting the waveband for a fit to the N~stars.  This procedure allowed
the estimation of two further values: $\sigma_{2,N}$ against $\sigma_{2,J}$.
Fig.~\ref{jfind_b} shows the ratio $\sigma_{1,J}$/$\sigma_{1,N}$ plotted 
against $\sigma_{2,J}$/$\sigma_{2,N}$ on logarithmic scales.  High positive 
numbers indicate a far better match of the star to the J~template whereas 
high negative numbers are found when the star is better matched with the 
N-star template.   The stars fall 
into two groups broadly following their initial classification, showing 
a good separation between N and J~stars.  It is clear that
using two measures gives a better isolation of the J~stars than can be
achieved with just one axis (or spectral feature).   It should also be noted
that the C$_2$-weak stars lie to the right-hand edge of the main body of 
non-J~stars showing that they can be distinguished from the typical cool 
N~stars through the shape of the CN bands in the 6200-6300\AA \/ wavelength 
range.  The dominant cause of the overlap between the two types is the
simplistic nature of the initial classification.   The few stars which
are seen in the border area between the J~stars and the rest were checked 
individually and classified accordingly.  The presence of stars in that 
area is mainly due to greater noise in their spectra.
The set of open circles represents the sample of J~stars used in this paper.

A set of stars associated with the main body of stars in 
Fig.~\ref{jfind_b} was chosen to represent non-J stars for use as 
a comparison with the J~stars.

\begin{figure}
\vspace{9.0cm}
\includegraphics{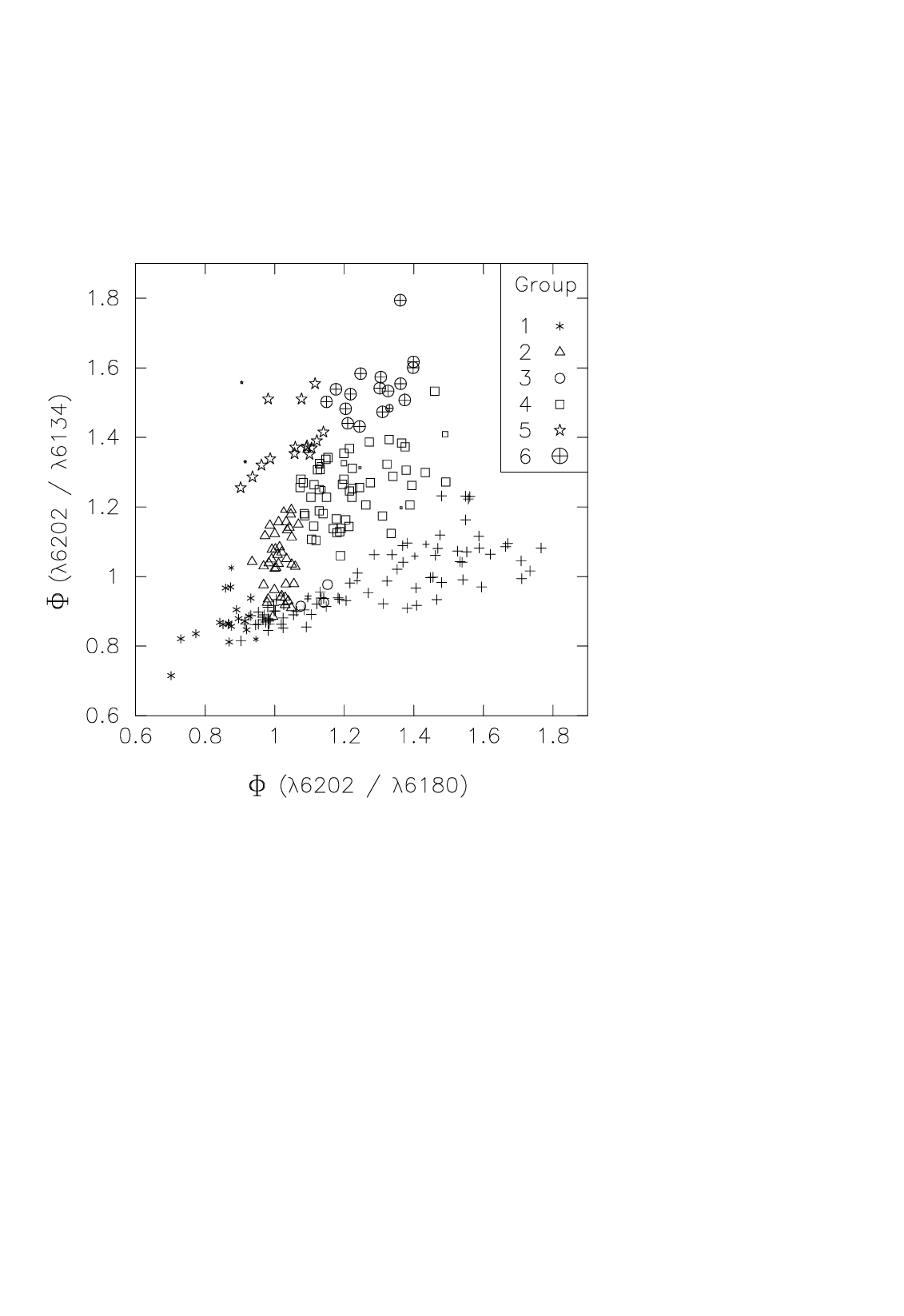}
\caption{Spectrum slopes.  The J-star spectral groups are as marked in the
inset.  The non-J comparison stars are denoted by plus signs.  Small 
symbols in this and later figures are for stars with weak spectra.}
\label{slopes}
\end{figure}
 
\section{J-Type Spectral Groups}

A pseudo-continuum was defined as two linear segments between maxima 
at 5722\AA, 6202\AA \/ and 6620\AA.  These are points where C$_2$ and 
CN band absorption is weakest.  Ideally, a point at 6750\AA \/ would 
be used (see Westerlund et al. \shortcite{westerlund91}) but the 
2dF spectra do not extend this far into the red.  For some stars, the 
maximum at 6180\AA \/ was greater than the one at 6202\AA, so the 
pseudo-continuum was defined by that point instead.  Minima associated 
with the principal lines and bands were then measured, as were maxima 
each side of these minima.  Band strengths were calculated with respect 
to the pseudo-continuum and also to a local continuum defined by the 
nearby maxima.  All this was done after applying a slight smoothing to 
the spectra using a 3-pixel box which is equivalent to the instrumental
resolution. 

Three important maxima lie immediately to the red of the heads of the 
(0,\,2) and (1,\,3) $^{12}{\rmn C}^{12}{\rmn C}$ bands and the  
(0,\,2) $^{13}{\rmn C}^{12}{\rmn C}$ band, at 6202\AA, 6134\AA \/ 
and 6180\AA \/ respectively.  These are marked by arrows on Fig.~\ref{plex}.  
Fig.~\ref{slopes} shows the relationship between two 
spectral slopes relating these maxima, $\Phi\,(\lambda6202/\lambda6134)$ 
and $\Phi\,(\lambda6202/\lambda6180)$, after normalizing to a 
flat pseudo-continuum.  This normalisation is particularly useful here
in order to remove instrumental effects in the 2dF spectra, due to errors 
in the CCD flat fielding and calibration procedures, and to chromatic 
effects inherent in the 2dF optics (see Lewis et al. \shortcite{lewis02} 
for details).  However, the corrections were relatively small here since 
the carbon star spectra are in the red spectral region and cover only 
a limited wavelength range.  

\begin{figure*}
\vspace{18.0cm}
\includegraphics{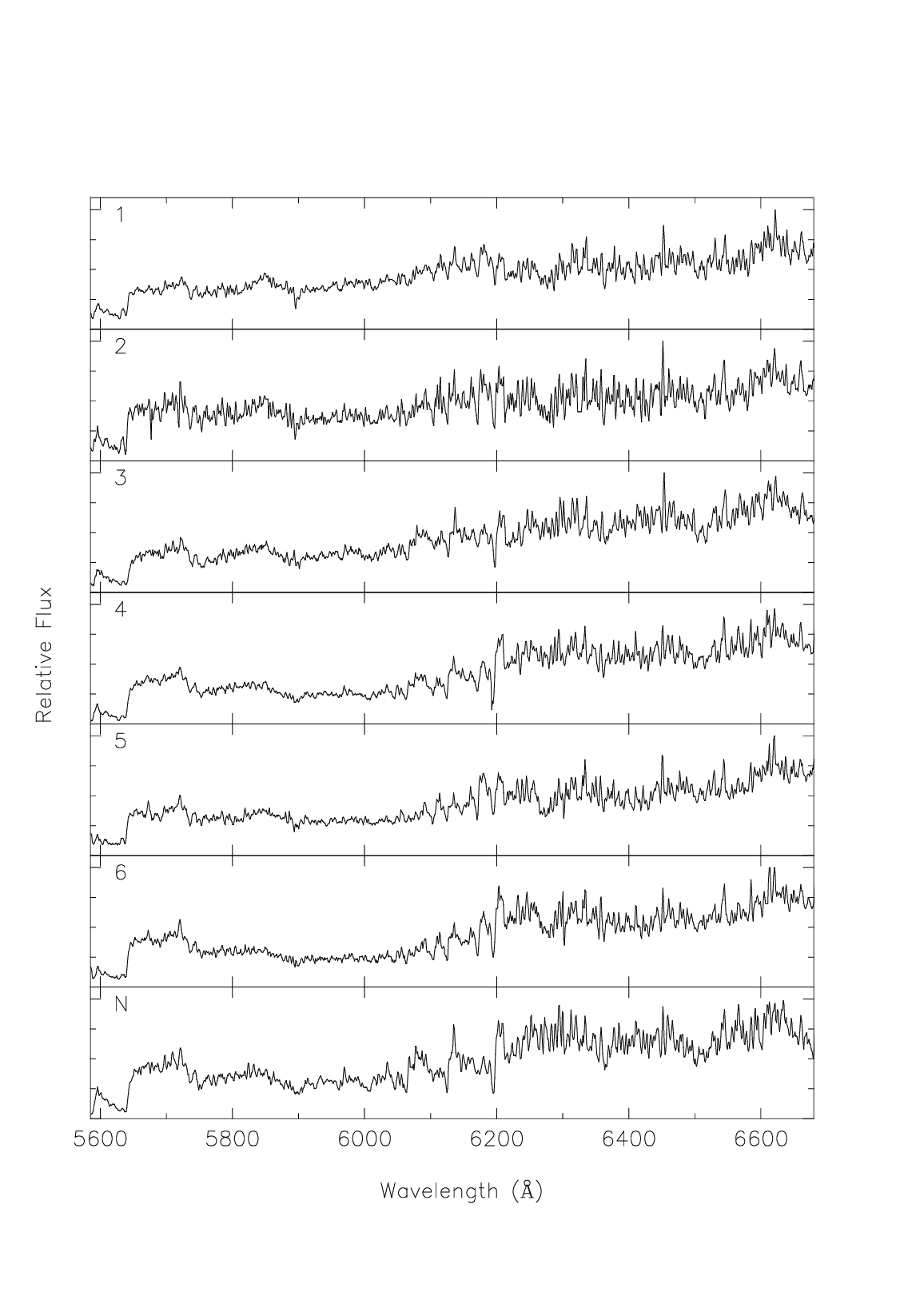}
\caption{Representative 2dF spectra in Groups 1-6: \/ \mbox{1\,--\,KDM\,5391,}
\mbox{2\,--\,KDM\,3643,} \mbox{3\,--\,KDM\,3391,}  
\mbox{4\,--\,KDM\,1410,} \mbox{5\,--\,KDM\,2622,} 
\mbox{6\,--\,KDM\,5544}.  A typical N-star (KDM\,3181) is shown as a 
comparison.  The plots are the reduced spectra after wavelength 
calibration and sky subtraction but with no intensity or flux calibration 
(for details see Cannon et al. in preparation); they are scaled to their 
maxima near 6600\AA \/ with the lower horizontal axis of each 
panel marking the zero flux level.  Portions of the lowest two spectra 
were used in Fig.~\ref{plex}.}
\label{sample}
\end{figure*}

It transpires that spectra within areas of this plot appear broadly 
similar and six broad groupings can be distinguished.  These are identified 
by different symbols in Fig.~\ref{slopes} amd some subsequent plots.  
Typical examples of each group are shown in Fig.~\ref{sample}.  
These arbitrarily defined groups are maintained to aid the following 
discussion.  Fig.~\ref{slopes} also shows the set of N-type comparison 
stars.  These lie below the J~stars in the plot, though there is some 
overlap, particularly towards the lower left-hand corner.
 
It is easy to see the key features of these groups by inspecting the
spectra in Fig.~\ref{sample}. (These are unsmoothed spectra.)
The 6134\AA \/ peak is stronger than the 6202\AA \/ peak in Group~1, 
these two peaks are at about the same level in Groups~2--3, whilst in
Groups~4--6 the 6134\AA \/ peak is below the other.
Spectral distinction within Groups~2--3 and Groups~4--6 is made by the 
level of the flux at the peak at 6180\AA \/ relative to the 
6134\AA \/ and 6202 \AA \/ peaks.  The best examples of these groups in the 
Galactic J-star spectra plotted by Barnbaum et al. \shortcite{barnbaum96} are
Group~1 -- WX~Cyg [J6~(C3$^-$ J4)] (though WX~Cyg is peculiar in some 
respects), Group~2 -- HO~Cas [J4.5~(C5 J5$^-$)], and Group~6 --
T~Lyr [J4:p~(C5 J3.5)].  The terms in the square brackets are the 
MK classifications based on the revised sheme of Keenan \shortcite{keenan93} 
as quoted by Barnbaum et al. \shortcite{barnbaum96}.  There are no 
representatives of Groups~3--5.

\section{Photometry}

Near-Infrared (NIR) photometry of sources in much of the LMC is 
available through the Second Incremental Data Release of the 
Two Micron All Sky Survey 
-- \mbox{2-MASS} \cite{skrutskie97} which is accessible on-line at 
http://irsa.ipac.caltech.edu.   \mbox{2-MASS} $JHK$ measurements were 
extracted for most of the carbon stars by matching catalogue coordinates.
Only two J~stars had ambiguous identifications in the \mbox{2-MASS} 
database.  The  $(J-K)$ colour is related to the effective temperature --
$T_{eff}$ (Bessell et al. \shortcite{bessell83}, 
Bergeat, Knapik \& Rutily \shortcite{bergeat01}).  

$IJK$ photometry is also available from the Deep Near-Infrared Survey 
of the southern sky -- DENIS \cite{cioni00}
which is available on-line from the CDS at Strasbourg 
(http://cdsweb.u-strasbg.fr/denis.html).  
The $J$ and $K$ data were extracted for the J~stars and used for those 
stars with no \mbox{2-MASS} measurements.  These magnitudes 
were made fainter by 0.15\,mag and 0.17\,mag respectively to bring them 
into alignment with the \mbox{2-MASS} data.  These corrections were  
calculated from the differences seen for the sample of J~stars measured in 
both surveys.  There was no magnitude dependence and the rms values
were 0.13\,mag and 0.10\,mag.
 
$RI$ photometry of the LMC stars is available from KDMK01.  It was derived 
from measurements of direct UKST plates made with the SuperCOSMOS 
measuring machine; see KDMK01 for details.  The $RI$ photometry in general 
suffers more from the blending of the carbon star images with close 
line-of-sight companions than the infrared photometry because the 
neighbouring stars are usually bluer than the carbon stars.  Nevertheless, 
the extraction software used to obtain the $RI$ photometry copes with 
blended images in all but the severest situations.  Also, a few stars 
suffer from contamination of $R$ due to H$\alpha$ emission from background
nebulosity. 

\begin{figure}
\vspace{8.5cm}
\includegraphics{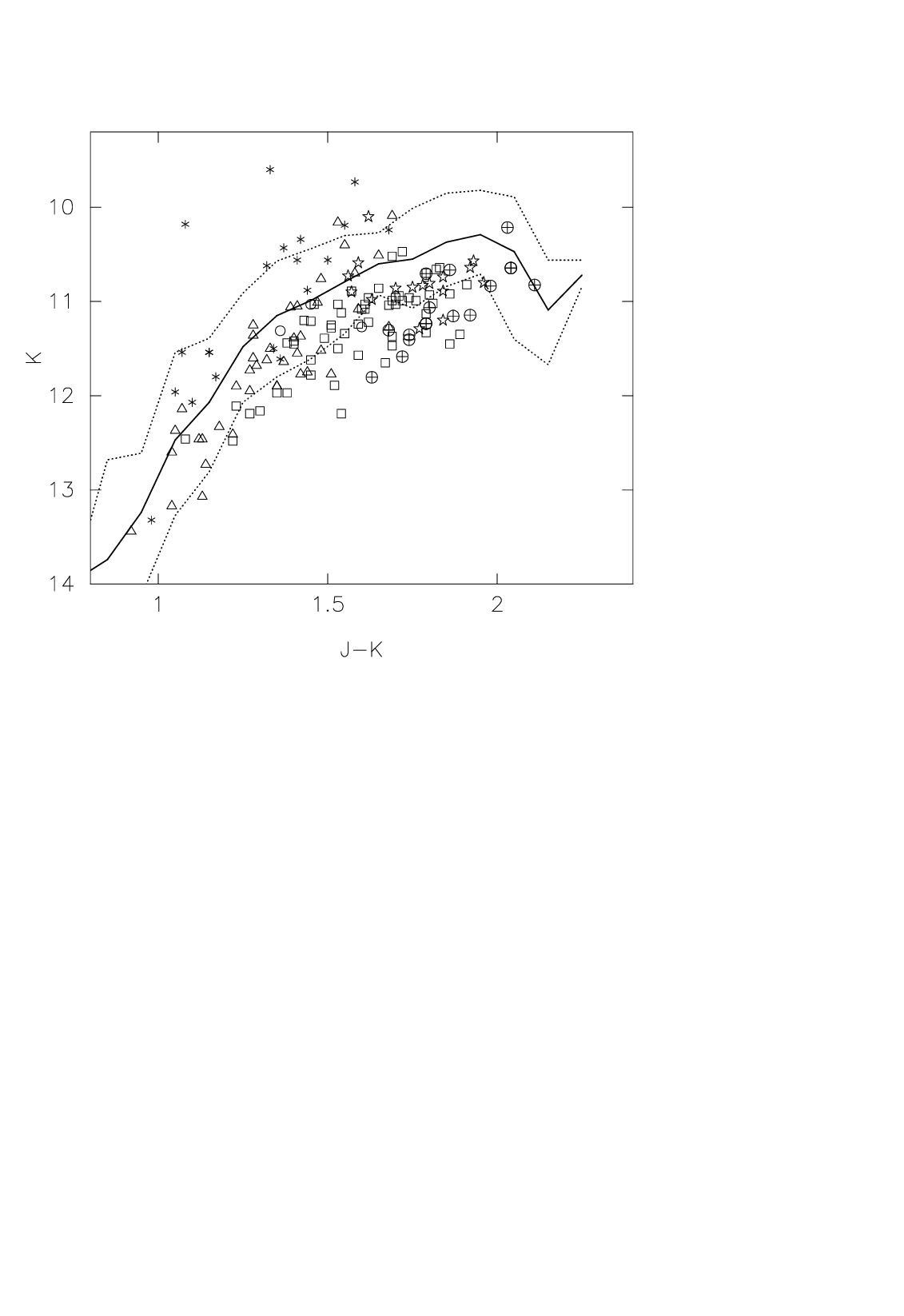}
\caption{$K$ vs $(J-K)$ colour-magnitude diagram. The symbols are as in 
Fig.~\ref{slopes}.  The solid line is the mean relation for N~stars in
the LMC and the dotted lines indicate the range containing
80 per cent of the population of N~stars.}
\label{jk_k}
\end{figure}

\begin{figure}
\vspace{8.5cm}
\includegraphics{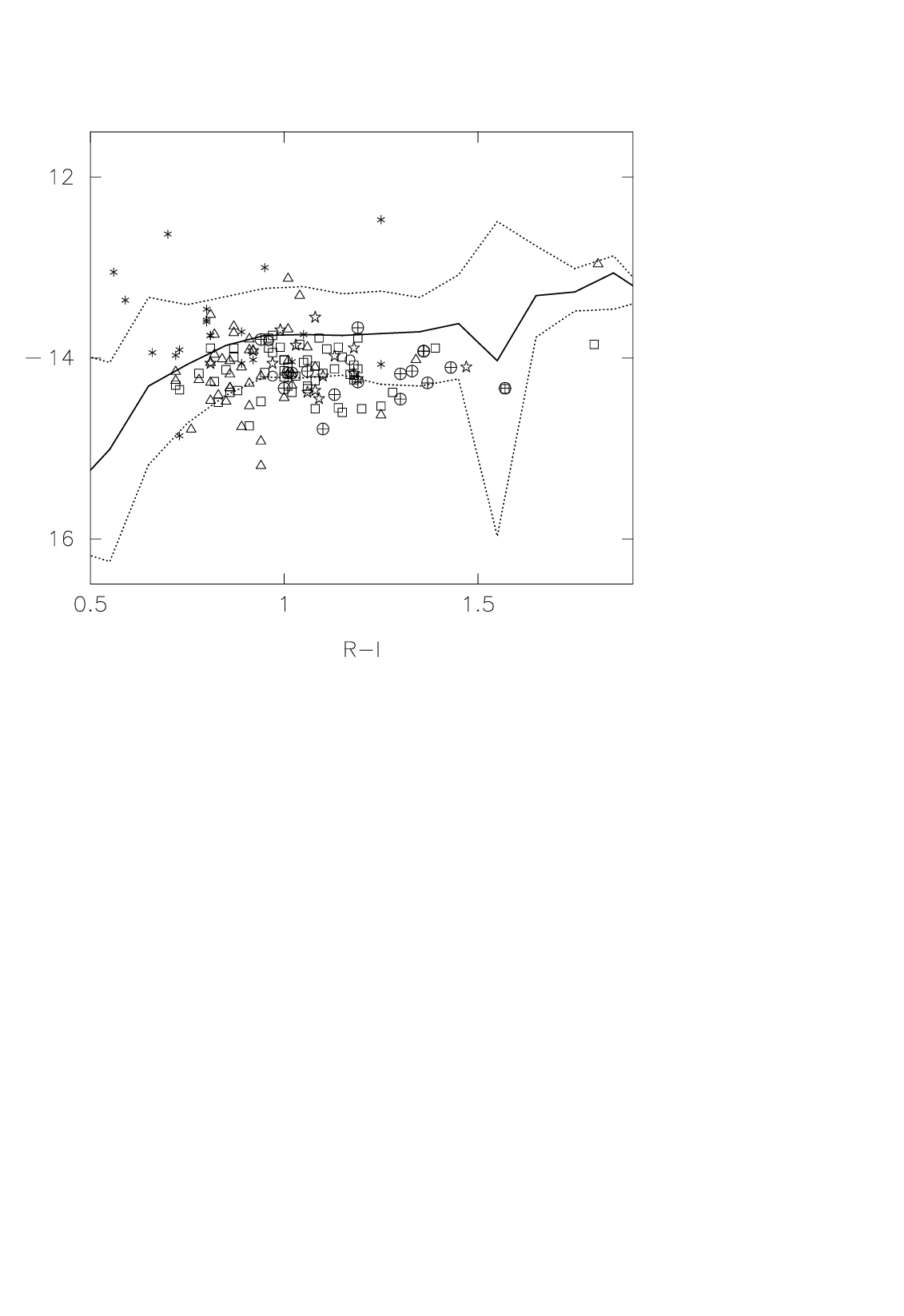}
\caption{$I$ vs $(R-I)$ colour-magnitude diagram. The symbols are as in 
Fig.~\ref{slopes}.  The solid line is the mean relation for N~stars in
the LMC and the dotted lines indicate the range containing
80 per cent of the population of N~stars.}
\label{ri_i}
\end{figure}

\begin{figure}
\vspace{8.5cm}
\includegraphics{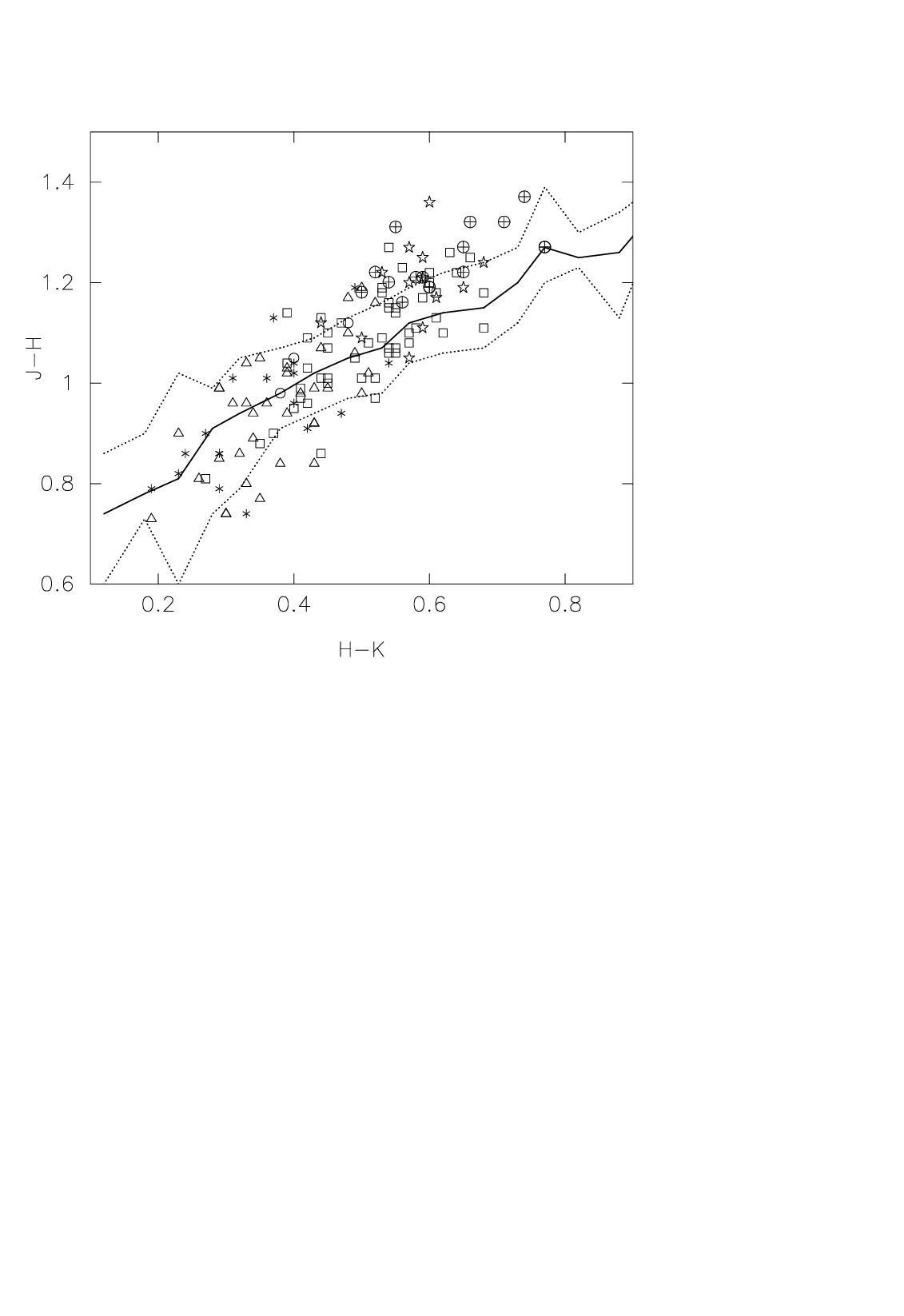}
\caption{$(J-H)$ vs $(H-K)$ two-colour diagram. The symbols are as in 
Fig.~\ref{slopes}.  The solid line is the mean relation for N~stars in
the LMC and the dotted lines indicate the range containing
80 per cent of the population of N~stars.}
\label{hk_jh}
\end{figure}

An infrared $K$, $(J-K)$ colour-magnitude diagram is shown 
in Fig.~\ref{jk_k}, again with the spectral groups as defined in 
Fig.~\ref{slopes} indicated.  Superimposed are lines 
representing the sequence of LMC N~stars (including the C$_2$-weak stars).  
This sequence was obtained from the $JHK$ photometry of the N~stars in the
2dF sample by contructing histograms in $K$ for each 0.1 mag interval
in $(J-K)$.  The solid line is the mean and the dotted lines show 
the extents of the region containing 80 per cent of the stars in each bin.
A full description of the work on the N~stars will be presented elsewhere. 
What is clear is that most of the J~stars form a sequence which
lies about 0.6 mag fainter than the N-star sequence.  This is certainly
so for stars redder than $(J-K)$\,=\,1.2, but caution must be applied to this 
statement for the bluer stars which are much fewer in number.    There 
are also some exceptions which are much brighter.  These are almost all 
Group~1 stars; in fact, all but one of the Group~1 stars lie to the bright 
side of the mean N-star sequence.  The sequence of faint
blue stars is almost all from Group~2 and the reddest stars are 
predominantly from Group~6.  Thus, there is a clear link between optical 
spectral characteristics on the one hand and temperature and/or luminosity 
(as determined from the NIR photometry) on the other. 

The displacement of the J~stars from the bulk of the
N~stars in Fig.~\ref{jk_k} has been interpreted here as a change in 
luminosity, although the effect could also be due at least in part 
to changes in the spectra, caused by molecular bands in the near-IR.  
Only infrared spectroscopy can settle this but the luminosity 
interpretation is favoured since the shifts in $(J-K)$ colour would be 
very large for many stars, and because a similar effect is seen in 
the $I$--$(R-I)$ CMD (see Fig.~\ref{ri_i} below), where the loci are 
almost horizontal.

The possible existence of two groups of J~stars with different luminosities
arising from stars of different initial mass and perhaps following different
evolutionary mechanisms has already been suggested (Lorenz-Martins 
\shortcite{lorenz96}, Abia \& Isern \shortcite{abia00}).  The photometric 
separation of Group~1 stars from the rest lends support to this idea.     

It will be useful in Section~6 to consider the dependence of various spectral 
parameters on the near-infrared photometry.  Normal J~stars will be taken 
to be those located in the broad band of stars centred on a sequence roughly 
parallel to but fainter than that of the N~stars.  Bright J~stars are 
taken to be those more than 0.6\,mag brighter in $K$ than this sequence.
Three colour ranges will prove to be useful: $(J-K) \leq 1.25$, 
$1.25 \leq (J-K) \leq 1.5$ and  $(J-K) \geq 1.5$. 

Some of the carbon stars are likely to be variable, so the photometry 
has this additional uncertainty.  However, with both infrared and visual 
observations having been almost always taken over short periods of time 
(see KDMK01 for details of the latter), stellar variations with typical 
periods of the order of several hundred days will not introduce large 
errors into either the $JHK$ or $RI$ colours.

Fig.~\ref{ri_i} is is the $I$--$(R-I)$ colour-magnitude diagram with the
N-star sequence shown.  There is less spread in colour or magnitude than in
Fig.~\ref{jk_k}, partly because the spectroscopy was carried out in the
$R$-waveband and introduces some bias.  Nevertheless, the J~stars are 
typically fainter than the N~stars in $I$ as they were in $K$ and the 
Group~1 stars are generally brighter than others.  

Fig.~\ref{hk_jh} is the equivalent near-infrared two-colour diagram. 
The slope of the J-star sequence is clearly steeper than that of the 
N-star sequence. i.e., the reddest J~stars have more flux in $H$ than 
N~stars of the same temperature ($(J-K)$ colour) and the opposite is true
for the bluest J~stars. The Group~1 stars which are noteworthy in 
Fig.~\ref{jk_k} as being bright do not distinguish themselves from 
the J-star sequences in this two-colour diagram.  

The effect of interstellar extinction is minimal.  The relevant 
extinction and colour excess per unit E($B-V$) in the LMC estimated 
from OB stars \cite{morgan82} are A$_K$\,=\,0.3, A$_I$\,=\,1.5,
E($J-K$)\,=\,0.7 and E($R-I$)\,=\,0.75.  Reddening measures, including 
foreground reddening, are typically E($B-V$)\,=\,0.1 \cite{westerlund97}
so such levels would have little 
effect on Fig.~\ref{jk_k}.  Larger extinction can be found in some 
associations, but this would not affect many of the carbon stars.
Hence, there is no obvious group of reddened stars in either Fig.~\ref{jk_k}
or Fig.~\ref{ri_i} (i.e. stars displaced to the right and slightly downwards).

Although the J and N~stars follow different sequences in all these diagrams, 
the breadths of the sequences are too great to allow any photometric method 
of distinguishing between J and N~stars.

\section{Spectral Analysis}

There are three main parameters in the classification of 
J~stars \cite{keenan93}: temperature, c-index (carbon index) and j-index
(isotopic index).  
The first is derived from atomic lines seen in the blue and cannot be 
directly applied to the bulk of 
the 2dF data.  Moreover, it is particularly difficult for faint 
carbon stars because the emitted flux in the blue is very weak. 
However, the infrared colour can be used as a temperature
indicator (Bessell et al. \shortcite{bessell83}, 
Bergeat et al. \shortcite{bergeat01}).  The c-index depends on the 
strong (0,\,0)
and (0,\,1) Swan bands at 5165\AA \/ and 5635\AA. The first of these is 
out of the 2dF spectral coverage and the second is difficult to determine 
because the flux on the blue side is weak and lies close to the edge of 
the detector.  (Also, the Jan 1998 data do not include this band.) 
Nevertheless, there are many other C$_2$ features that can be used for this 
purpose.  The j-index is easier to determine because it is based on
several bands which all lie within the wavelength range covered by the
2dF spectra.  These indices 
and the other spectral features described in this section of the paper 
were measured from the spectra after being smoothed with a 3-pixel box,
matching the resolution of the instrument.

\subsection{j-index}
A J~star was for many years defined as one for which 
the strength of the $^{13}{\rmn C}^{12}{\rmn C}~\lambda6168$ band was 
at least half that of the $^{12}{\rmn C}^{12}{\rmn C}~\lambda6122$
band \cite{gordon68}.  The definition in the revised classification scheme 
for carbon stars \cite{keenan93} is slightly different in that there
the j-index is based on the
ratios of the bands $^{13}{\rmn C}^{12}{\rmn C}~\lambda6168$ to
$^{12}{\rmn C}^{12}{\rmn C}~\lambda6192$, 
$^{13}{\rmn C}^{12}{\rmn C}~\lambda6102$ to
$^{12}{\rmn C}^{12}{\rmn C}~\lambda6122$ and
$^{13}{\rmn C}^{14}{\rmn N}~\lambda6260$ to
$^{12}{\rmn C}^{14}{\rmn N}~\lambda6206$.

\begin{figure}
\vspace{8.0cm}
\includegraphics{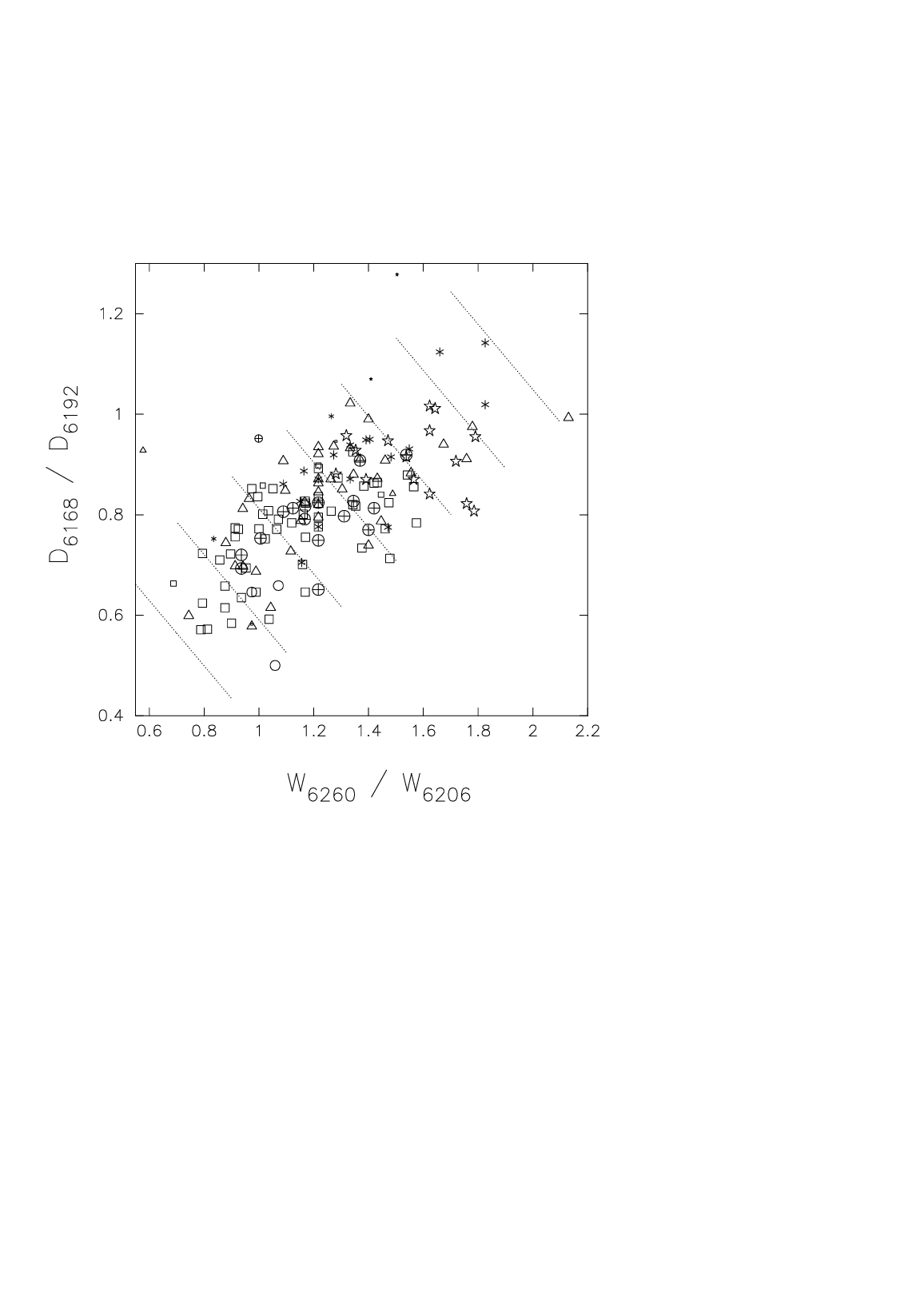}
\caption{Definition of the j-index.  The ratio of the depths of the 
(0,\,2) $^{13}{\rmn C}^{12}{\rmn C}~\lambda6168$ 
and $^{12}{\rmn C}^{12}{\rmn C}~\lambda6192$ bands is plotted against 
the ratio of the equivalent widths of the 
(4,\,0) $^{13}{\rmn C}^{14}{\rmn N}~\lambda6260$ and 
$^{12}{\rmn C}^{14}{\rmn N}~\lambda6206$ bands. 
The symbols are the spectral groups as in Fig.~\ref{slopes} and the 
dotted lines indicate from 
left to right the positions of the j-indices 3.5, 4.0 ..... 6.5.}
\label{jindex}
\end{figure}

The main difficulty in determining the j-index of a carbon star, and 
indeed other spectral parameters, is that defining the continuum is not 
straightforward.  The local continuum and 
pseudo-continuum are greatly different in Group~6 stars for instance,
as can be seen in Fig.~\ref{sample}.
Clearly, the huge depression of the local continuum is not due to 
$^{13}{\rmn C}^{12}{\rmn C}$, so the local continuum is more appropriate
for defining the j-index.  Two measures are used in this paper to define 
the j-index.  The first is the ratio
of the (0,\,2) bands $^{13}{\rmn C}^{12}{\rmn C}~\lambda6168$ and 
$^{12}{\rmn C}^{12}{\rmn C}~\lambda6192$ measured from the local continuum 
to the minimum (single pixel) of the band.  The ratio of the 
equivalent widths (W$_\lambda$) of the (4,\,0) $^{12}{\rmn C}^{14}{\rmn N}$
and $^{13}{\rmn C}^{14}{\rmn N}$ bands provides the second measure.  
W$_{6260}$ and W$_{6206}$ were determined from the template 
fitting procedure described in Section 3, using the mean J-star template.  
Fig.~\ref{jindex} shows the ratio of C$_2$ band depths 
D$_{6168}$\,/\,D$_{6192}$ plotted against the ratio W$_{6260}$\,/\,W$_{6206}$.
The two ratios are well correlated, in fact better than similar 
ratios based on other continuum levels.  So a j-index was defined as 
the distance along the sequence as marked in the figure.  It was chosen to be
similar to the MK index \cite{keenan93} which runs from 3.5 to 7 in a J~star 
\cite{barnbaum96}.   The stars with the lowest j-indices have 
$^{13}{\rmn C}^{12}{\rmn C}~\lambda6168$ bands slightly stronger than
0.5 times their $^{12}{\rmn C}^{12}{\rmn C}~\lambda6122$ bands as is 
expected from the older definition of the J~star \cite{gordon68}.  There 
is a slight divergence at high j-index levels between Group~1 and 
Group~5 stars which is probably due to difficulties in defining the 
continuum levels for such different types of spectrum.

\begin{figure}
\vspace{9.0cm}
\includegraphics{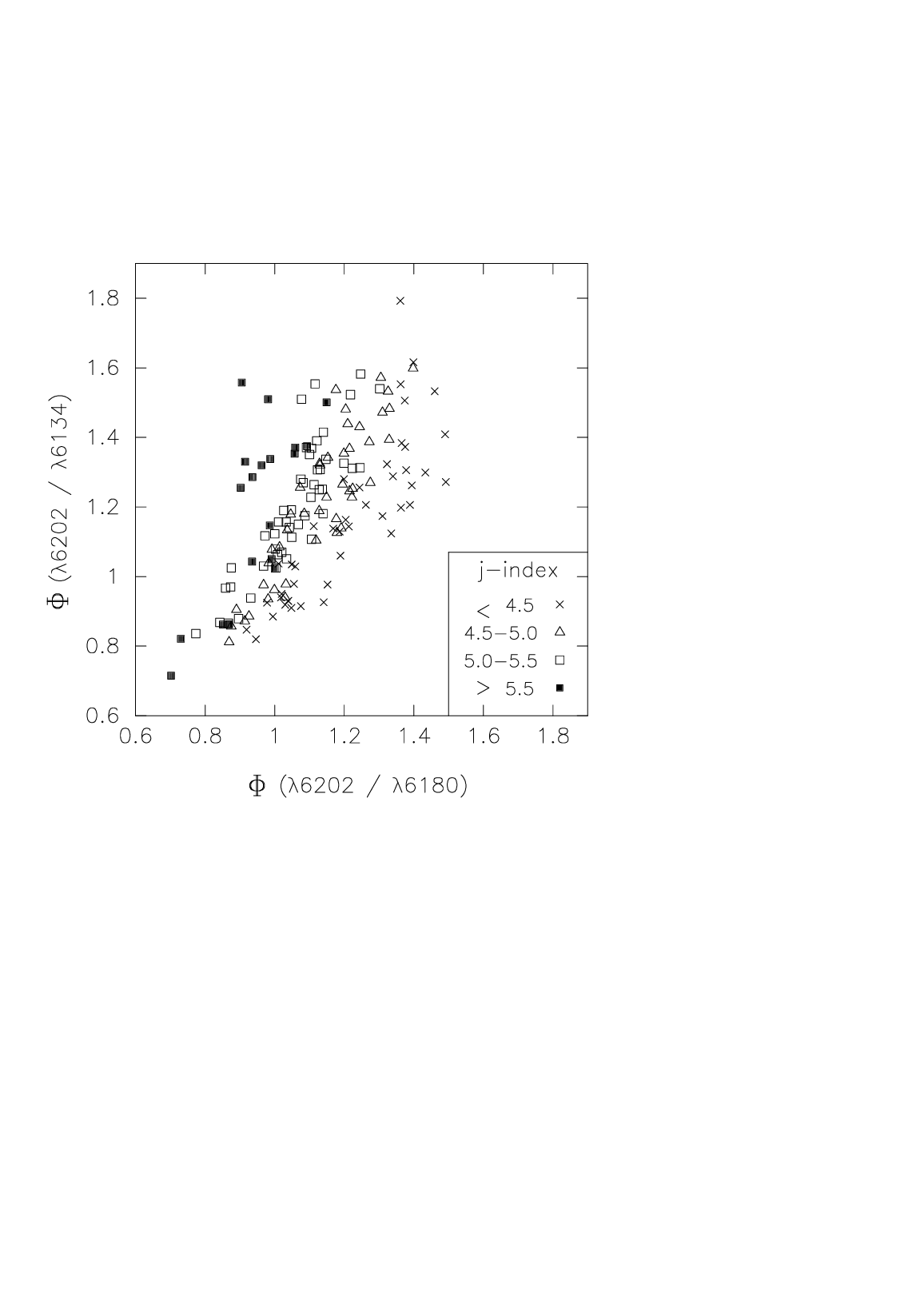}
\caption{Fig.~\ref{slopes} with the symbols marking ranges of j-index as
shown in the inset.}
\label{slopes_j}
\end{figure}

Fig.~\ref{slopes_j} is Fig.~\ref{slopes} redrawn with symbols representing
the j-index as determined from Fig.~\ref{jindex}.  It is clear that these
spectral parameters are controlled by the j-index, with zones of constant
j-index roughly given by sectors defined by straight lines through a point 
near the lower left-hand corner of the plot, with the j-index increasing 
in the anti-clockwise direction.  As is to be expected, observational 
errors and probably real fluctuations in these parameters give rise to 
the considerable overlap between the different symbols.  No discrimination 
is possible in the lower left-hand corner where the Group~1 stars are found.

\subsection{c-index}

\begin{figure}
\vspace{8.0cm}
\includegraphics{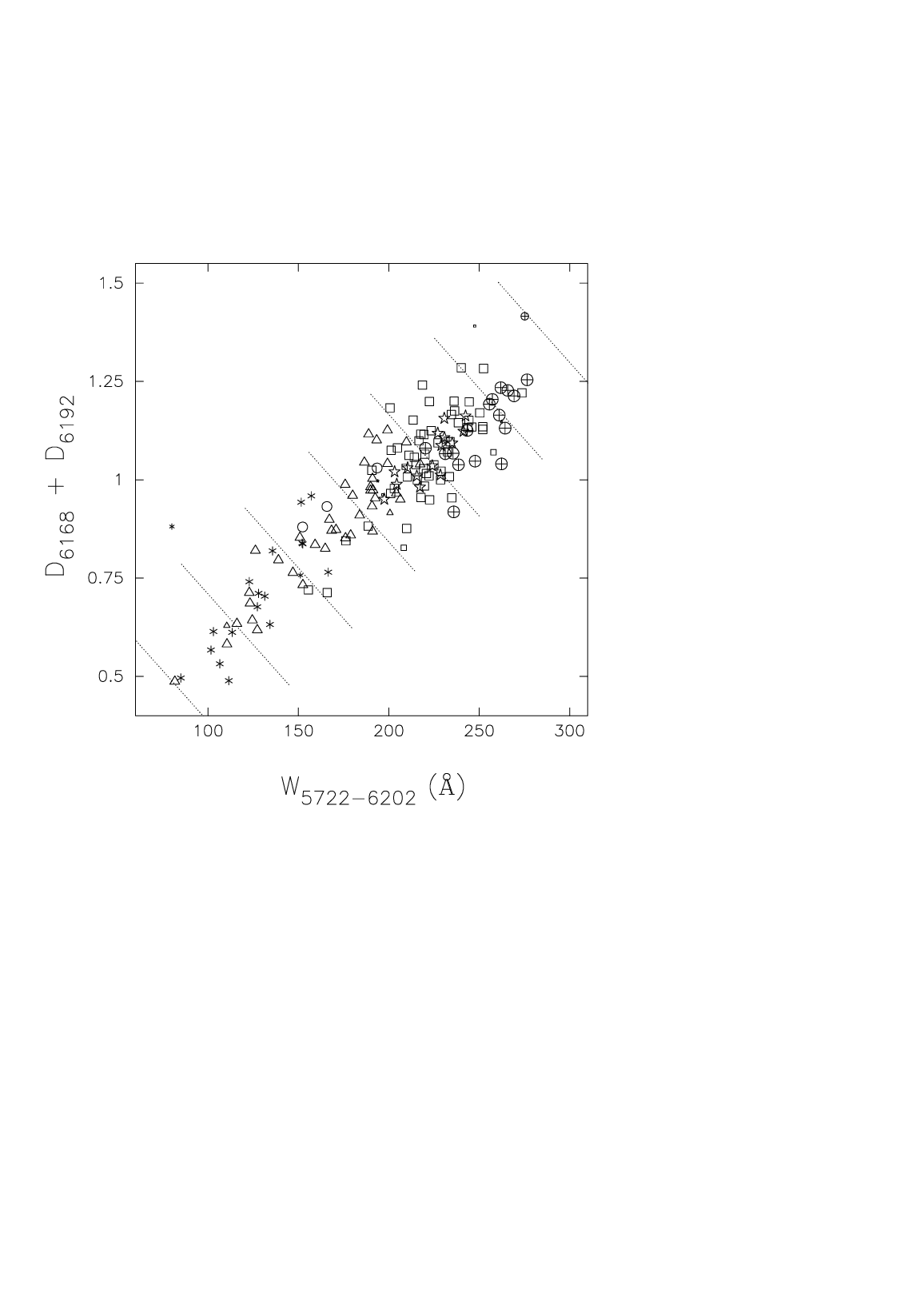}
\caption{Definition of the c-index.  The sum of the depths of the
(0,\,2) $^{13}{\rmn C}^{12}{\rmn C}~\lambda6168$ 
and $^{12}{\rmn C}^{12}{\rmn C}~\lambda6192$ bands is plotted against 
the equivalent width of the entire spectrum from 5722\AA \/ to 6202\AA \/. 
The symbols identify the six spectral groups as in Fig.~\ref{slopes} and 
the dotted lines indicate from left to right the positions of the c-indices 
3.5, 4.0 ..... 6.5.}
\label{c_index}
\end{figure}

The carbon strength can be treated in a similar way. For J~stars, carbon 
needs to be measured from both isotopic bands rather than just one when
the results would clearly be modified by the j-index.  Thus, one measure of
carbon strength is the sum of the (0,\,2) 
$^{13}{\rmn C}^{12}{\rmn C}~\lambda6168$ \/ and 
$^{12}{\rmn C}^{12}{\rmn C}~\lambda6192$ bands.  Another is the 
complete absorption between
5722\AA \/ and 6202\AA, whatever its precise cause.  The equivalent width 
of this (W$_{5722-6202}$)
was measured relative to the pseudo-continuum.  Fig.~\ref{c_index} 
shows the relationship between these two measures.  The correlation is 
obvious and there is no lateral dependence on j-index.  Hence, a c-index 
was set as the position along this sequence as indicated in the figure.   

\begin{figure}
\vspace{9.0cm}
\includegraphics{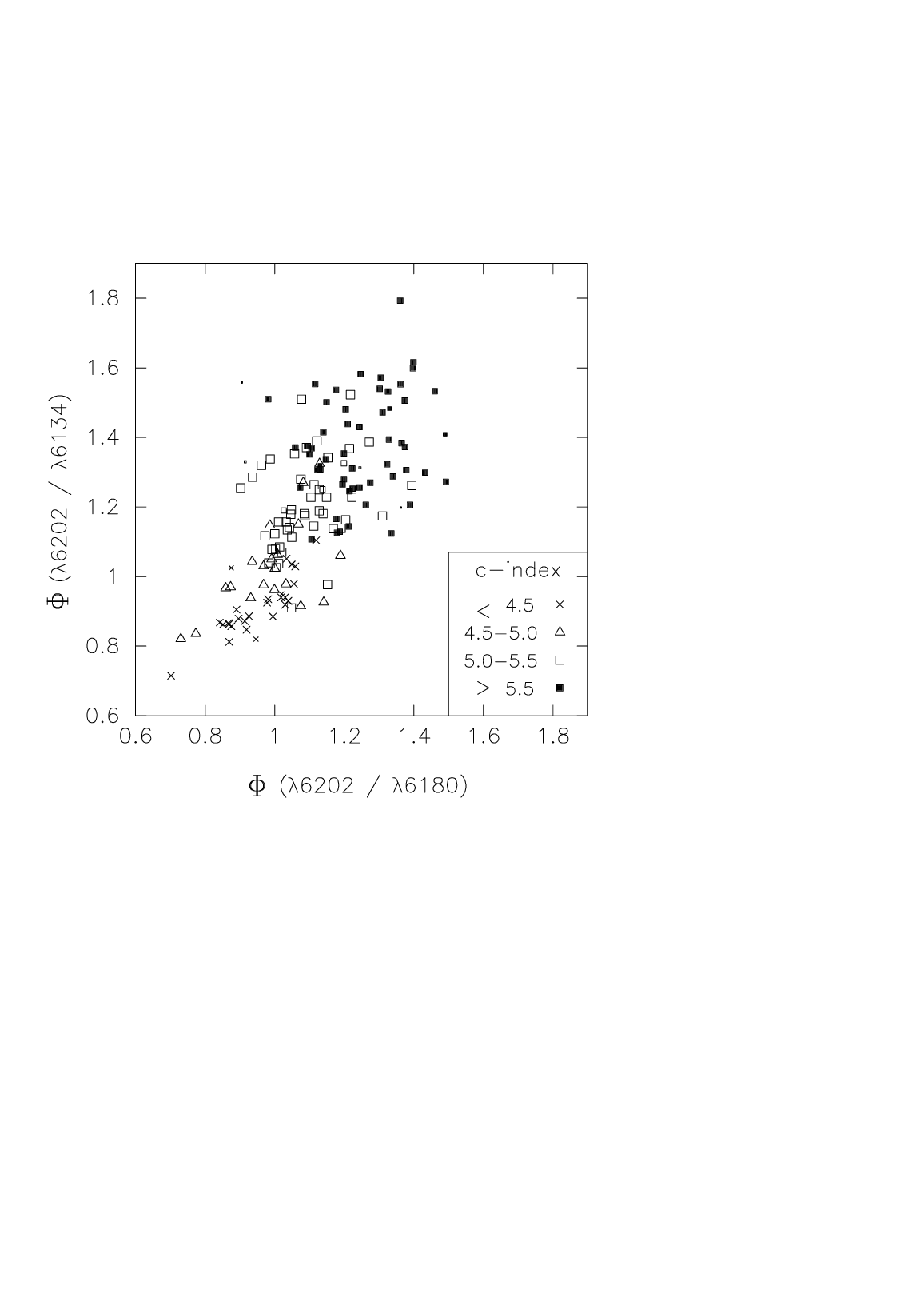}
\caption{As Fig.~\ref{slopes} with the symbols marking ranges of c-index
as shown in the inset.}
\label{slopes_c}
\end{figure}

It transpires that the spectral groups are dependent on both the c-index 
and the j-index as can be seen in Fig.~\ref{slopes_c} which is 
Fig.~\ref{slopes} redrawn this time with symbols 
showing the c-index. In this case, the c-index increases 
clockwise in sectors of a circle centred near the point (1.4,\,1.0).  
By comparing Figs~\ref{slopes_j} and \ref{slopes_c} it is clear that 
the two spectral slopes ($\Phi$) give a two-dimensional classification 
that can be related to the j- and c-indices and are more easily measured 
than the latter.

\begin{figure}
\vspace{8.0cm}
\includegraphics{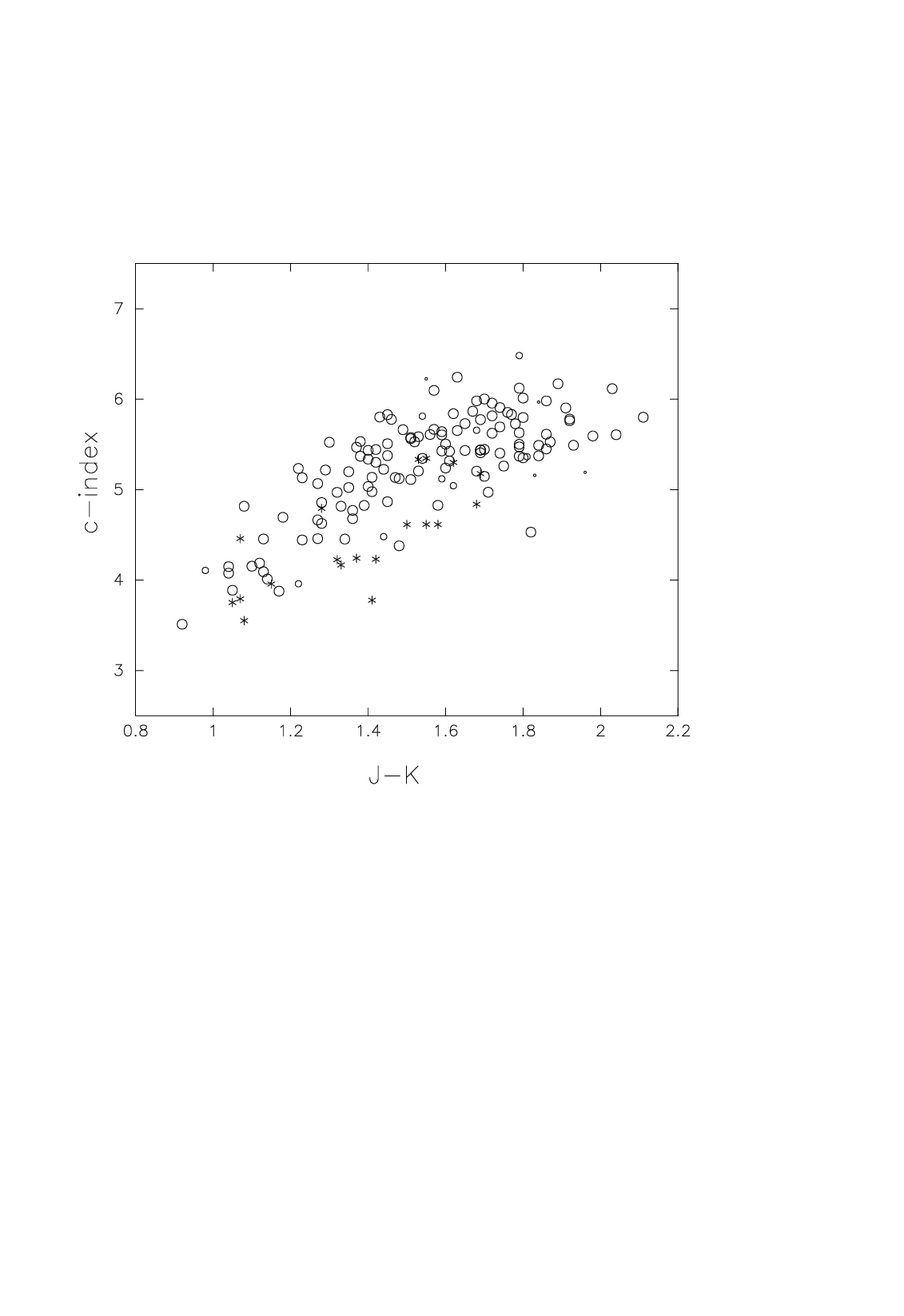}
\caption{C-index against $(J-K)$.  The circles represent the normal
J~stars and the asterisks represent the bright J~stars.}
\label{c_t}
\end{figure}

Fig.~\ref{c_t} shows the c-index plotted against $(J-K)$ (temperature).  
The trend is as expected; the cooler stars have the higher c-indices.  
It should be noted that the index here is purely a measure of the observed 
carbon strength and not a measure of that strength relative to a standard 
of the same temperature.  There is considerable spread, but that is 
significantly reduced when the bright J~stars are excluded.
There is a distinct tendency for the brighter J~stars to have weaker carbon
bands (smaller c-indices) than the fainter (normal) J~stars of the same
$(J-K)$ colours.  

\subsection{C$_2$ bands}

\begin{figure}
\vspace{8.0cm}
\includegraphics{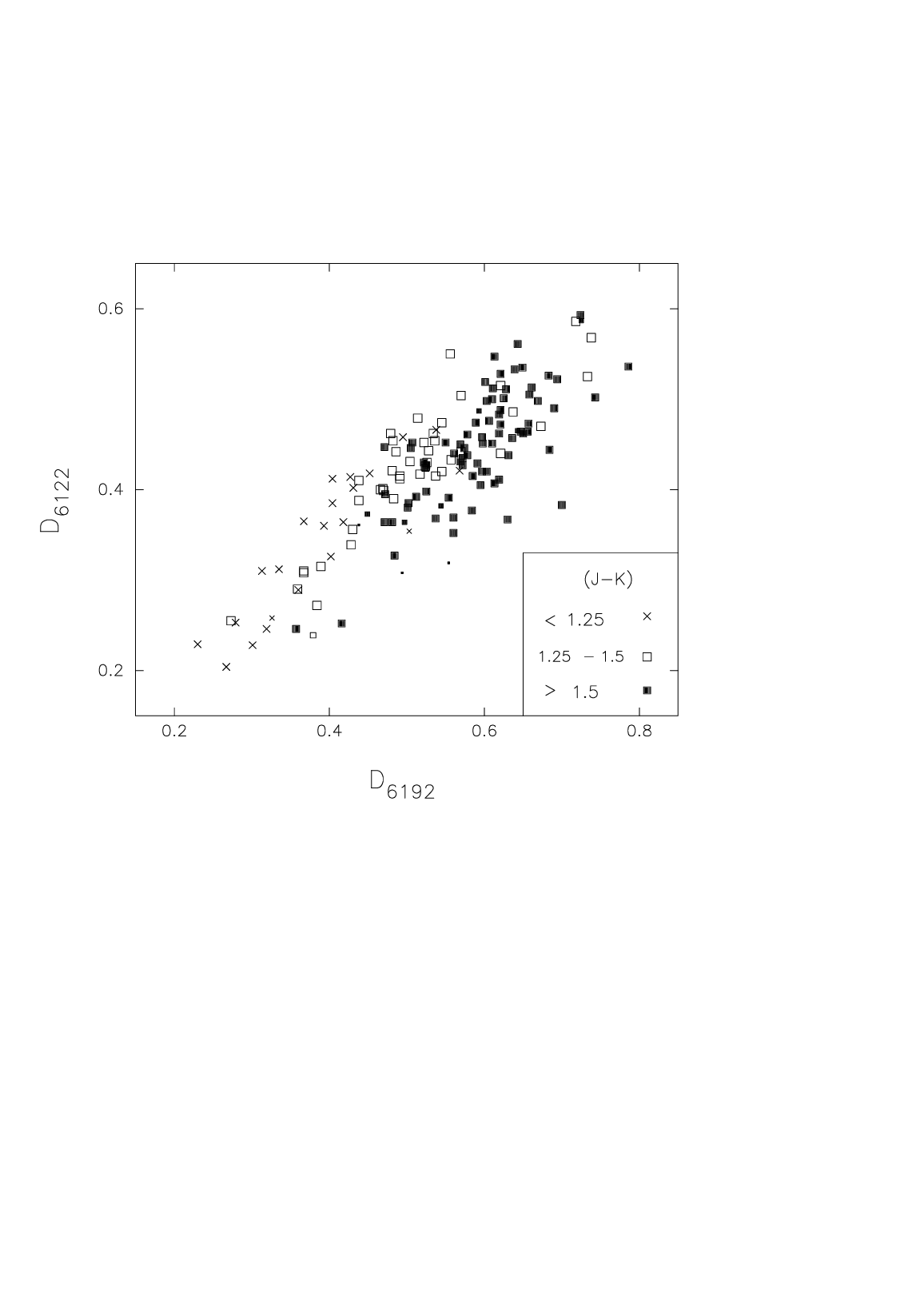}
\caption{Strengths of the C$_2$ bands at 6122\AA \/ and 6192\AA. 
The symbols mark ranges of $(J-K)$.}
\label{c_c}
\end{figure}

Fig.~\ref{c_c} shows the expected correlation between the depths of the
$^{12}{\rmn C}^{12}{\rmn C}$ (1,\,3) and (0,\,2) bands at 6122\AA \/ and 
6192\AA \/ respectively.  The symbols mark the $(J-K)$ colour.  It is 
clear that the relationship between D$_{6122}$ and D$_{6192}$
is different for each colour range in that it appears progressively to 
the right as colour increases (i.e. temperature decreases).   This would 
give rise to a correlation between the ratio of the band
depths, D$_{6122}$\,/\,D$_{6192}$, and $(J-K)$ in the sense that the 
(1,\,3) band is stronger with respect to the (0,\,2) band for the bluer 
(hotter) stars.  This ratio changes from 0.9 at $(J-K)\sim 1$
to 0.6 at $(J-K)\sim 2$ with a spread of $\sim\pm 0.1$.  It is tempting 
to think that this is indeed the effect of temperature increasing the 
strength of the higher energy band.  However, it is more likely that it 
is caused by the continual change in shape of the local continuum which 
occurs as the large absorption trough between 5720\AA \/ and 6200\AA \/ 
increases with overall carbon strength, which in turn changes with the 
temperature.

The strong (0,\,1) $^{12}{\rmn C}^{12}{\rmn C}~\lambda5635$ band is only 
available for those stars observed during the November 1998 run and so
is of limited use for this work.  However, it was measured as an
equivalent width between the spectral peak at 5590\AA \/ and the
top of the band head near 5650\AA \/ using the latter point as the 
continuum level.  There is a clear correlation between this band and 
both parameters used above to determine the c-index.  

\begin{figure}
\vspace{8.0cm}
\includegraphics{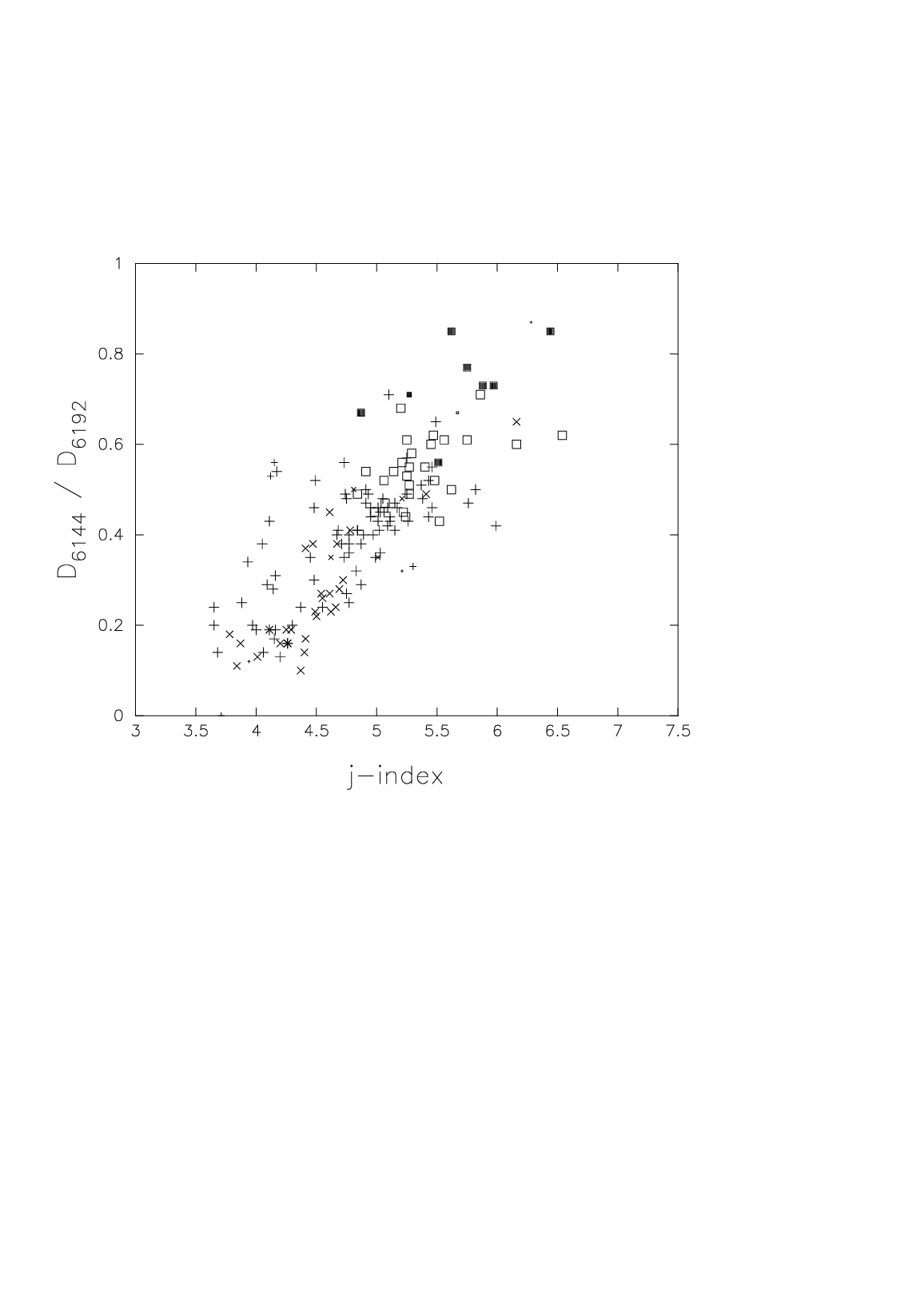}
\caption{Ratio of the depths of the bands 
$^{13}{\rmn C}^{13}{\rmn C}~\lambda6144$ to
$^{12}{\rmn C}^{12}{\rmn C}~\lambda6192$ against 
j-index. The symbols are: plus -- no $^{13}{\rmn C}^{13}{\rmn C}~\lambda6144$ 
obvious; open square -- moderate $^{13}{\rmn C}^{13}{\rmn C}~\lambda6144$;
filled square -- strong $^{13}{\rmn C}^{13}{\rmn C}~\lambda6144$.  
In addition, the crosses mark those stars where any detected feature 
is slightly offset from the expected wavelength for 
$^{13}{\rmn C}^{13}{\rmn C}~\lambda6144$.}
\label{c13c13}
\end{figure}

When the $^{13}$C abundance is very high, the 
(0,\,2) $^{13}$C$^{13}$C\,$\lambda$6144
bandhead is often seen alongside the other (0,\,2) bandheads at 6168\AA \/ 
and 6192\AA.  Fig.~\ref{c13c13} shows that this band strengthens as
the j-index increases, as is to be expected.  However, there
can also be minima in this region due to the various components
of the (9,\,4) $^{12}{\rmn C}^{14}{\rmn N}$ band and, for border-line
J~stars, the Ba\,{\sc ii}\,$\lambda$6141 line.  The J~star spectra were 
inspected by eye and assigned a code according to whether 
$^{13}{\rmn C}^{13}{\rmn C}~\lambda6144$ was strong, moderate or
undetectable.  As expected, those stars with 
$^{13}{\rmn C}^{13}{\rmn C}~\lambda6144$ identified by eye are all in 
the upper-right part of the diagram and shown as open or filled squares.
However, they are accompanied by a significant number of stars for 
which no feature had been identified by eye.  Stars in the last category 
are shown as a plus sign in Fig.~\ref{c13c13}.  Some shown as crosses
have measured features displaced slightly redwards of the expected
position of the $^{13}{\rmn C}^{13}{\rmn C}~\lambda6144$ band and are 
clearly due to another cause.  Nevertheless, there remains a correlation
in the figure for the plus signs.  Inspection of the spectra for stars
with moderately high j-indices and where no obvious 
$^{13}{\rmn C}^{13}{\rmn C}~\lambda6144$ band is visible shows that
the band would be located near the centre of a broad depression of the 
spectrum between peaks near 6130\AA \/ and 6160\AA \/ (see the high 
resolution spectrum of Y~CVn plotted by Barnbaum \shortcite{barnbaum94}).  
From Fig.~\ref{c13c13} it would seem that this depression is due to 
$^{13}{\rmn C}$, and the absence of an obvious 
$^{13}{\rmn C}^{13}{\rmn C}~\lambda6144$ band within it for so many 
stars would suggest that the ordinate is measuring a combined feature.

\subsection{CN bands}

\begin{figure}
\vspace{8.0cm}
\includegraphics{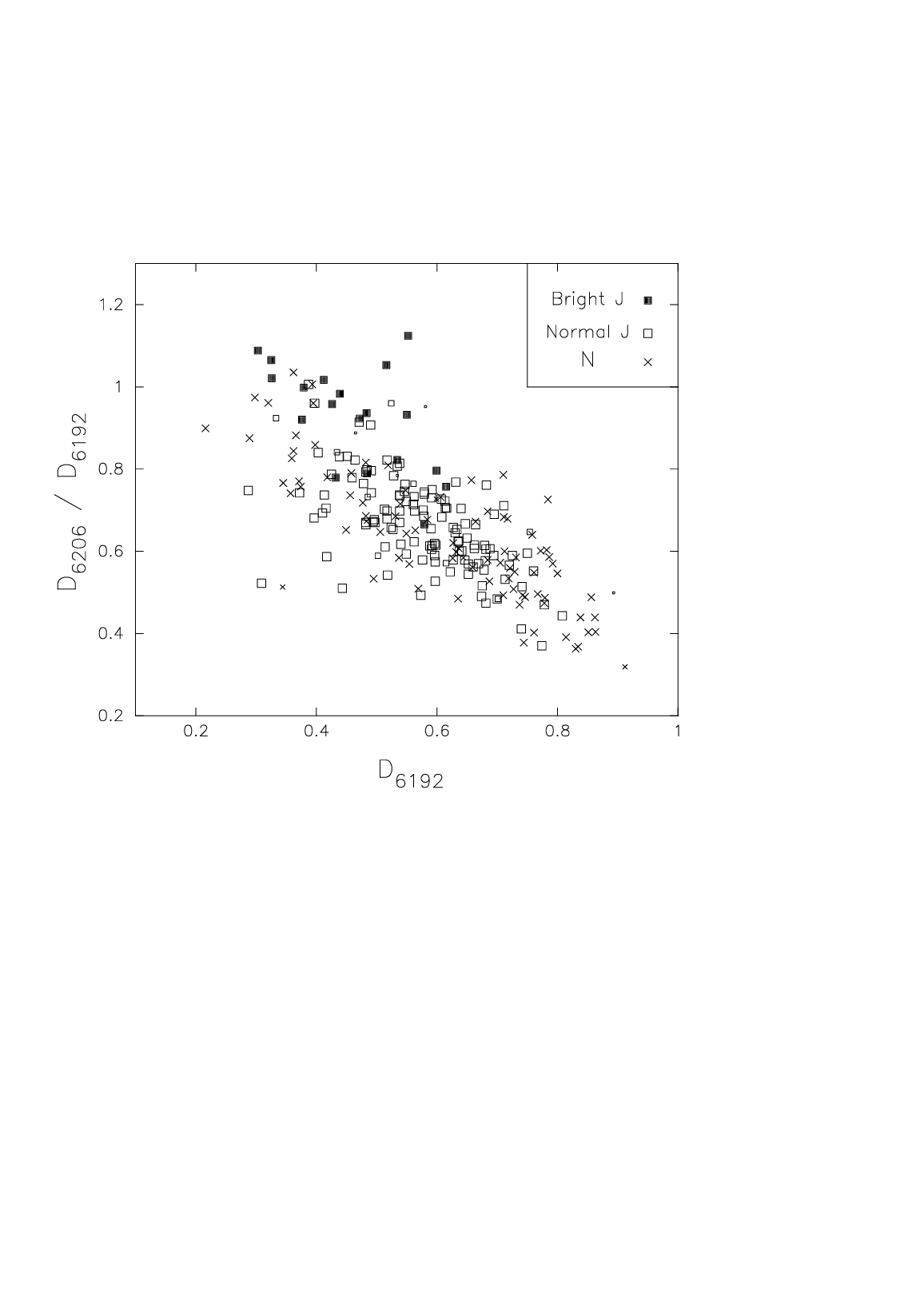}
\caption{Ratio of the depths of $^{12}{\rmn C}^{14}{\rmn N}~\lambda6206$ to 
$^{12}{\rmn C}^{12}{\rmn C}~\lambda6192$ plotted against the depth of 
$^{12}{\rmn C}^{12}{\rmn C}~\lambda6192$.  Bright J~stars are marked 
by filled squares and normal J~stars by open squares.  The comparison 
N~stars are shown by the crosses.}
\label{cn_c}
\end{figure}

Fig.~\ref{cn_c} shows the ratio of the depths of the 
$^{12}{\rmn C}^{14}{\rmn N}~\lambda6206$ and 
$^{12}{\rmn C}^{12}{\rmn C}~\lambda6192$ bands 
plotted against the $^{12}{\rmn C}^{12}{\rmn C}~\lambda6192$ band 
depth with symbols identifying the bright and faint J~stars.  Most of 
the normal J~stars lie along a broad sequence which shows a slight decrease 
in the strength of $^{12}{\rmn C}^{14}{\rmn N}~\lambda6206$
relative to $^{12}{\rmn C}^{12}{\rmn C}~\lambda6192$ as 
$^{12}{\rmn C}^{12}{\rmn C}~\lambda6192$ increases.  A similar sequence
is seen for the comparison N~stars.   More obvious is the observation that
the bright J~stars consistently have stronger 
$^{12}{\rmn C}^{14}{\rmn N}~\lambda6206$ bands than the fainter (more common) 
stars with the same $^{12}{\rmn C}^{12}{\rmn C}~\lambda6192$ band depth.
This is in line with the statements by Keenan \shortcite{keenan93} that 
the CN bands show a positive luminosity effect.  However, it may not be 
simply the effect of surface gravity but may be associated with the 
chemical evolution of these bright J~stars.

Part of the explanation of the varying 
$^{12}{\rmn C}^{14}{\rmn N}/^{12}{\rmn C}^{12}{\rmn C}$ ratio might lie 
in the effect of the (9,\,4) $^{13}{\rmn C}^{14}{\rmn N}$ band.
Its R1 and Q1 bandheads lie at 6197\AA \/ and 6202\AA, coinciding with
the maximum between the $^{12}{\rmn C}^{12}{\rmn C}~\lambda6192$ and 
$^{12}{\rmn C}^{14}{\rmn N}~\lambda6206$ bands.  So, increasing the 
j-index reduces the local maximum which will cause a greater reduction 
of the weaker band - the CN band, thus reducing the ratio 
$^{12}{\rmn C}^{14}{\rmn N}/^{12}{\rmn C}^{12}{\rmn C}$.  On the
other hand, the band has more effect redwards of this point than 
towards the blue \cite{marenin72}.  However, the (9,\,4) band is 
relatively weak and probably not the cause of what is seen.  

\subsection{Na D Lines}

The Na D lines have been used as a temperature index in the past, but 
this is no longer considered reliable mainly because they fall in the 
middle of the broad absorption region between 5720\AA \/ and 6200\AA \/
which can significantly affect their appearance.  Nevertheless, they 
could still provide useful information about the stars.  One problem
with fibre spectroscopy is the difficulty of sky subtraction for
strong sky lines such as the Na~D lines.  However, the radial
velocity of the LMC has the effect of shifting the Na~D lines arising 
from the LMC so that the D2 line is blended with the D1 line from 
the sky and the D1 line is clear of both Na~D sky lines.

The Na\,{\sc i}\,$\lambda$5896 line was measured in all the 
spectra by first fitting a polynomial to the region around 5900\AA \/ and 
secondly by fitting a gaussian profile to all features that diverged 
from that continuum.  Any feature (line) found, after correcting for
the LMC radial velocity, within 1.5\AA \/ of 5895.5\AA \/ was taken to 
be Na\,{\sc i}\,$\lambda$5896.  The equivalent width of 
Na\,{\sc i}\,$\lambda$5896 -- W$_{5896}$ -- was taken to be the equivalent 
width of the gaussian fit less any contribution from 
Na\,{\sc i}\,$\lambda$5890 in cases when the two lines did overlap. 

Measurements of W$_{5896}$ range from 0.0\AA \/ to 1.6\AA \/ with a mean 
and rms of 0.7\,$\pm$\,0.4\,\AA.  There is one J~star -- KDM\,2832
-- with a stronger Na~D line of W$_{5896}$\,=\,2.4\AA.  This star also
shows the strongest Li\,{\sc i}\,$\lambda$6708 line in the sample (see 
Morgan et al., in preparation), and in that sense at least is similar to
the cool J~star WX~Cyg \cite{barnbaum96}.  The sample of 
comparison N~stars have slightly stronger Na~D lines with a mean and rms of 
1.2\,$\pm$\,0.6\,\AA. 

\begin{figure}
\vspace{8.0cm}
\includegraphics{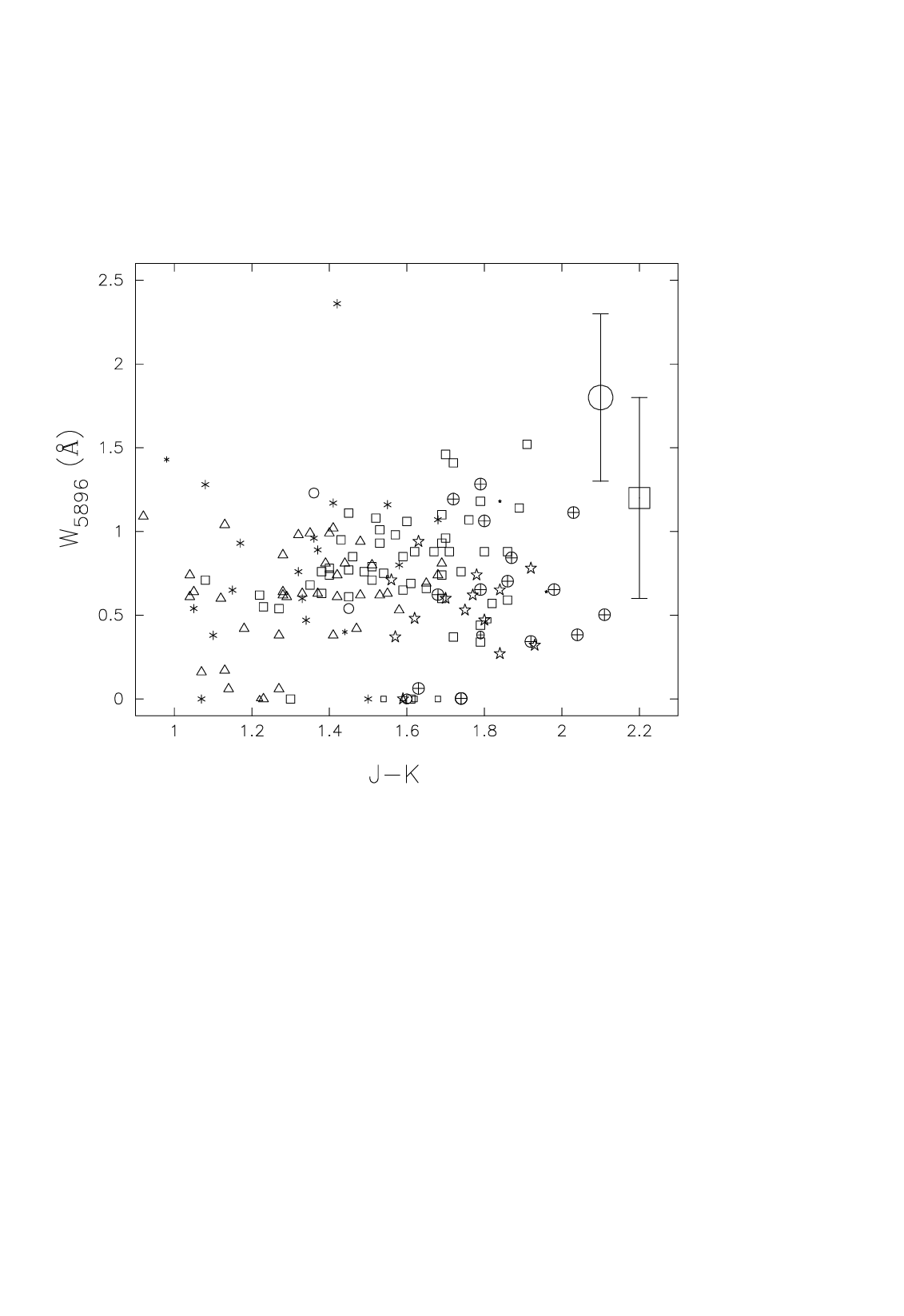}
\caption{W$_{5896}$ against ($J-K$).  The symbols represent the spectral 
groups as in Fig.~\ref{slopes}.  The large symbols shows the range of values
of W$_{5896}$ seen in the LMC comparison N~stars (square) and in 
Galactic J~stars (circle); the colour positions are arbitrary.}
\label{nad}
\end{figure}

W$_{5896}$ does not appear to depend strongly on ($J-K$), j-index, c-index or
spectral group.  Perhaps the bluest stars have slightly weaker Na~D lines
but the effect, if real, is small.  This is shown in Fig.~\ref{nad} for
($J-K$).  W$_{5896}$ is often set to zero when it is too small to be
measured; so the gap seen in Fig.~\ref{nad} at W$_{5896}\sim$0.2\AA \/
is not necessarily real.  The evidence certainly does not support the 
idea that the Na~D lines could be a good temperature indicator for LMC
J~stars.

It is interesting to compare W$_{5896}$ for LMC and Galactic stars.  The
latter were measured using {\sc iraf} from the spectra of stars presented by 
Barnbaum et al. \shortcite{barnbaum96}.  For the N~stars, only one (the
hottest) has W$_{5896}<2$\AA.  The Galactic J~stars have W$_{5896}$ mostly 
in the range $1.8\pm0.5$\AA \/ which is indicated in Fig.~\ref  {nad}.  
These levels are generally weaker than in Galactic N~stars as was noted 
above for the LMC stars.  The levels of W$_{5896}$ for the Galactic J~stars
are close to or greater than the upper limits seen in the LMC J~stars.  
W$_{5896}$ in some of the the Galactic R~stars is weaker again, the 
weakest being 0.9\AA.  Over half the values for the LMC stars of any type 
are smaller than this.  Thus, W$_{5896}$ is significantly weaker in the 
LMC than in the Galaxy.  This is a consequence of the lower metallicity
in the LMC, for Westerlund et al. \shortcite{westerlund95} have already 
shown that the Na~D lines are very much weaker in the metal-poor SMC 
than in the metal-rich Galactic Bulge.  None of the LMC J~stars has a 
Na~D line strong enough to warrant the specification of the NaD-index 
\cite{barnbaum96} in its classification.

\subsection{Blue spectral data}

\begin{figure}
\vspace{8.0cm}
\includegraphics{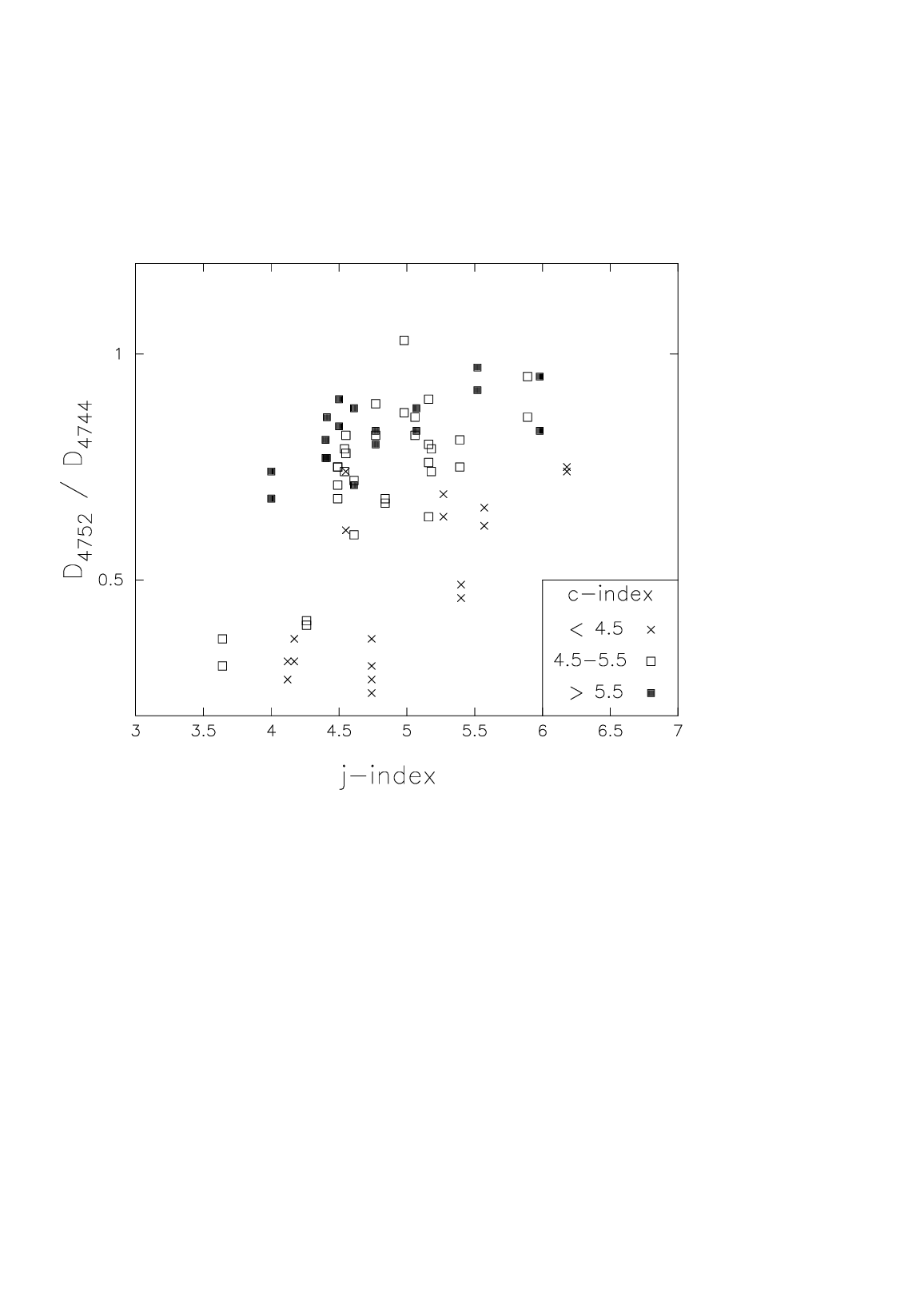}
\caption{Ratio of the depths of the $^{13}{\rmn C}^{13}{\rmn C}~\lambda4752$
and $^{13}{\rmn C}^{12}{\rmn C}~\lambda4744$ bands plotted against the 
j-index. The symbols mark the c-index as shown in the inset.}
\label{blue}
\end{figure}

The blue data from the 2002 service observations provided additional 
spectra of $\sim$300 stars in the wavelength range 4190--5290\AA.  These 
could, in principle, provide an alternative estimate of the j-index.
Fig.~\ref{blue} shows the ratio of the depths of the (1,\, 0) 
$^{13}{\rmn C}^{13}{\rmn C}~\lambda4752$ and
$^{13}{\rmn C}^{12}{\rmn C}~\lambda4744$ bands plotted against the j-index
with symbols indicating the c-index.  The depths were measured from a
continuum set near 4780\AA.  Each group of carbon stars plotted shows 
a trend of increasing $^{13}{\rmn C}$ (as measured by the ratio 
D$_{4752}$/D$_{4744}$) with j-index, but the ratio is systematically smaller
for the C-weak stars.  Thus, there is a discrepancy between the 
$^{13}{\rmn C}$ strengths measured from the blue bands and 
from the red bands (the j-index of this paper).  This discrepancy arises 
because the strong blue bands, both normal and isotopic, are often heavily 
saturated and therefore less sensitive to the $^{12}{\rmn C}^{13}{\rmn C}$ 
ratio, as was noted by Barnbaum et al. \shortcite{barnbaum96} in relation 
to the $^{12}{\rmn C}^{12}{\rmn C}~\lambda4737$ band.  This is far less 
of a problem for the weaker red bands.  The features of the plot are
consistent with this interpretation.

\subsection{Repeatability}

Seventeen stars were observed more than once.  The mean differences
between the indices obtained from pairs of spectra of the same star
are 0.17\,$\pm$\,0.14 for the j-index and 0.28\,$\pm$\,0.21 for the 
c-index.   These differences are $\sim$10 per cent of the overall 
spread of the indices which is $\simeq$\,3.5.  The errors are greater 
for the c-index because it is based on absolute line strengths and 
therefore more directly affected by the zero point errors that can 
arise in fibre-optic instruments. The j-index is based on line ratios.   
The spectral group classifications were the same in 14 of the 17 stars;
two of the three different classifications were for stars found very 
close to the group boundaries. 

\subsection{Tabulated results} 

A list of the J stars is given in Table~\ref{data}.  Stars observed more than 
once have a single entry based on the better quality spectrum.  
Column~1 is the name of the star in the KDMK01.
Columns 2 and 3 give the $RI$ the photometry taken from KDMK01
and Columns 4--6 give either the $JHK$ photometry 
resulting from the cross-identification with the 2-MASS Survey or the
$JK$ photometry from the DENIS Survey, adjusted as 
described in Section~5.
Column 7 is the spectral group number, Columns 8 and 9 are the 
c-index and j-index respectively, and Column 10 (labelled 
$^{13}{\rmn C}^{13}{\rmn C}$) is a code noting the
strength of the $^{13}{\rmn C}^{13}{\rmn C}~\lambda6144$ band (2 -- strong, 1
-- moderate, 0 -- undetected) as plotted in Fig.~\ref{c13c13}.
Finally, Column 12 marks some stars of special note:  those with a
`$\dagger$' and `$\ddagger$' identify stars with weaker spectra -- i.e.,
with counts at the 6200\AA \/ peak less than 600 and 300 respectively;
and the `*' identifies stars with a cross-indentification given in the 
footnotes: CRW -- Crabtree, Richer \& Westerlund \shortcite{crabtree76},
HC -- Hartwick \& Cowley \shortcite{hartwick88} and Cowley \& Hartwick 
\shortcite{cowley91}, WORC -- Westerlund et al. \shortcite{westerlund78}.
A photometric entry of `0.00' indicates that no measurement was available.   
`$H$\,=\,0.00' identifies DENIS $JK$ photometry.

\begin{figure*}
\vspace{18.0cm}
\includegraphics{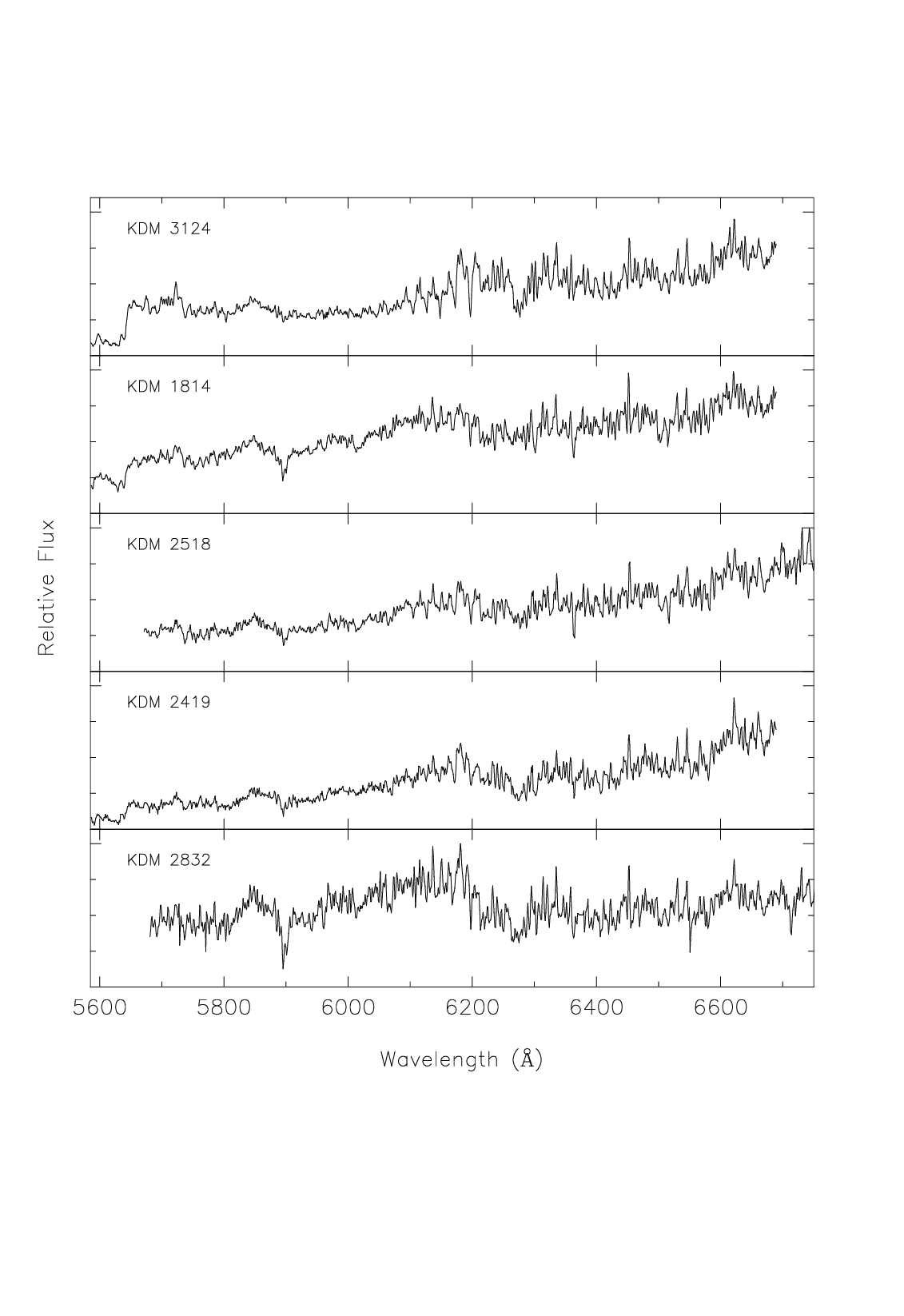}
\caption{2dF spectra of individual stars noted in the text.  The plots are 
the reduced spectra after wavelength calibration and sky subtraction but 
with no intensity or flux calibration; they are scaled to their 
maxima within the plotted wavelength range with the lower horizontal 
axis of each panel marking the zero flux level.}
\label{indspec}
\end{figure*}

\section{Individual stars} 

Three of the LMC J~stars have already been described in the literature; 
their 2dF spectra are plotted in Fig.~\ref{indspec} for comparison purposes.
Fig.~\ref{indspec} also shows the 2dF spectra of two other extreme J~stars.
 
\underline{KDM\,3124} \/ This is also WORC\,131 and its spectrum 
has been published as Star 4-9 by Crabtree et al. \shortcite{crabtree76}.
It is said to `exhibit the most extreme J-star characteristics of any 
star known to the authors'.  The 2dF spectrum has much higher resolution 
and S/N; points to note are: the extreme strength 
of the (4,\,0) $^{13}{\rmn C}^{14}{\rmn N}~\lambda6260$ band, the very 
strong (0,\,2) $^{13}{\rmn C}^{13}{\rmn C}~\lambda6144$ band, and the 
(0,\,1) $^{12}{\rmn C}^{12}{\rmn C}~\lambda5635$ band located as a minor 
feature on the `side' of the (0,\,1) $^{13}{\rmn C}^{12}{\rmn C}~\lambda5625$ 
bandhead.  Crabtree et al. \shortcite{crabtree76} quote a class 
\mbox{J4 (C4, J5+)} and Richer et al. \shortcite{richer79} quote 
\mbox{J4 (C5, J5+)}; the present work has KDM\,3124 as a Group~5 star -- 
\mbox{J (C5.3, J5.6)}.  No star in the 2dF sample has the 
$^{13}{\rmn C}^{13}{\rmn C}~\lambda6144$ band stronger than KDM\,3124, 
but a number of stars achieve j-indices that are just as high.  

\underline{KDM\,1814} \/ This star is Star 6-6 \cite{crabtree76} 
quoted as class \mbox{J4 (C4, J4+)}.  The present work has KDM\,1814 as 
a Group~1 star \mbox{J (C3.6, J4.8)}, i.e., a star with weak carbon.  
The difference could lie in the statement by Crabtree et al. 
\shortcite{crabtree76} that the (0,\,0) and 
(0,\,1) $^{12}{\rmn C}^{12}{\rmn C}$ bands are moderately strong, but the 
(0,\,2) $^{12}{\rmn C}^{12}{\rmn C}$ band and others in the 
$\Delta\nu$\,=\,+2 series are weak.  
The star is also WORC\,78 and its classification by Richer et al. 
\shortcite{richer79} is the same as that by Crabtree et al. 
\shortcite{crabtree76}.  It is also HC\,185 in the list of suspected LMC
CH~stars \cite{hartwick88}.  There is no evidence of significant 
photometric variation as seen in the infrared photometry from \mbox{2-MASS},
DENIS and Feast \& Whitelock \shortcite{feast92}, and the $I$ photometry
from DENIS and KDMK01.

\underline{KDM\,2518} \/  According to Richer et al. \shortcite{richer79}, 
this star which is WORC\,106 is \mbox{J4 (C4, J4)}.  The present work has 
\mbox{J (C4.2, J4.9)} for a Group~1 star.  KDM\,2518 is also 
HC\,160.  As for KDM\,1814, there is no evidence of 
significant photometric variation. 

\underline{KDM\,2419} \/  Like KDM\,3124, this star is an extreme j-star
with a j-index of 6.5 and a c-index of 4.6.  However, it belongs to spectral
Group~1 rather than Group~5 as KDM\,3124 described above.  The most striking
feature is the very large change in continuum gradient around the
peak near 6180\AA.    

\underline{KDM\,2832} \/  This star has the strongest Na~D lines in the 
2dF sample, and, as noted in Section~6.5, also has the strongest 
Li\,{\sc i}\,$\lambda$6708 line.  KDM\,2832 has a j-index of 5.6 and 
a c-index of 4.2, and, like the three stars plotted immediately above it 
in Fig.~\ref{indspec}, is a Group~1 star.

\section{Discussion} 

\subsection{Comparison with MK indices}

The same analysis as used here for the LMC J~stars can be carried out 
in a limited way on the eight J~star spectra published by 
Barnbaum et al. \shortcite{barnbaum96} and available from the CDS.  
Indices were adopted for these stars by comparing measured band ratios 
with those obtained for the LMC stars; in the case of the
c-index, W$_{5722-6202}$, W$_{5565-5645}$ and D$_{6192}$+D$_{6168}$,
and for the j-index, W$_{6260}$/W$_{6206}$, D$_{6168}$/D$_{6192}$,
and D$_{6168}$/D$_{6122}$.  The W$_{6260}$/W$_{6206}$ ratio was measured 
directly from the spectra rather than using the template fitting procedure.
Table~\ref{comp} provides a comparison of the indices measured here 
for these eight stars and the published MK indices.  Some spectral group 
types are borderline.  

\begin{table}
\begin{center}
\caption{Classification of selected Galactic J~stars}
\label{comp}
\begin{tabular}{lllc}
\hline
Star    & MK class         & Present class & Spectral group \\
        &                  &    &      \\
HD10636 & J4(C5.5, J5.5)   & J(C4.9, J4.9) & 2 \\ 
EU And  & J5$^-$(C5, J3.5) & J(C4.5, J4.6) & 1 (or 2) \\  
T Lyr   & J4(C5, J3.5)     & J(C5.6, J4.6) & 6 \\  
NQ Cas  & J4.5(C5, J4)     & J(C4.6, J4.2) & 2 \\  
HO Cas  & J4.5(C5, J5$^-$) & J(C4.5, J4.7) & 2 \\ 
HD19557 & J4(C5.5, J5.5)   & J(C4.9, J4.9) & 2 (or 5) \\  
VX And  & J4.5(C5, J5)     & J(C5.6, J3.9) & 6 \\  
WX Cyg  & J6(C3$^-$, J4)   & J(C3.7, J5.9) & 1 \\  
\hline
\end{tabular}
\end{center}
\end{table}

It is clear from Table~\ref{comp} that there is no obvious correlation 
between j-indices measured here and the published values.  The c-indices
do show some similarities in that a low value was found in both cases for 
WX~Cyg (the unusually strong Na~D absorption was excluded from 
W$_{5722-6202}$).  However, no such relationship exists for the other 
stars, though it should be remembered that the range in c-index covered 
by the sample is not large.  It was therefore 
impossible to make a transformation between the two sets of indices. 
The reasons for the differences in classification can
be seen by inspection of the published spectra.  For example, EU~And 
and T~Lyr are both classified as (C5,\,J3.5)
but have grossly different spectra.  Some differences could be
introduced by the fact that the MK c-indices were based on slightly 
different bands, but T~Lyr certainly shows a strong 
$^{13}{\rmn C}^{14}{\rmn N}$ band at 6260\AA, not a weak one as 
predicted by the J3.5 classifier.

\subsection{Indices}

\begin{figure}
\vspace{14.0cm}
\includegraphics{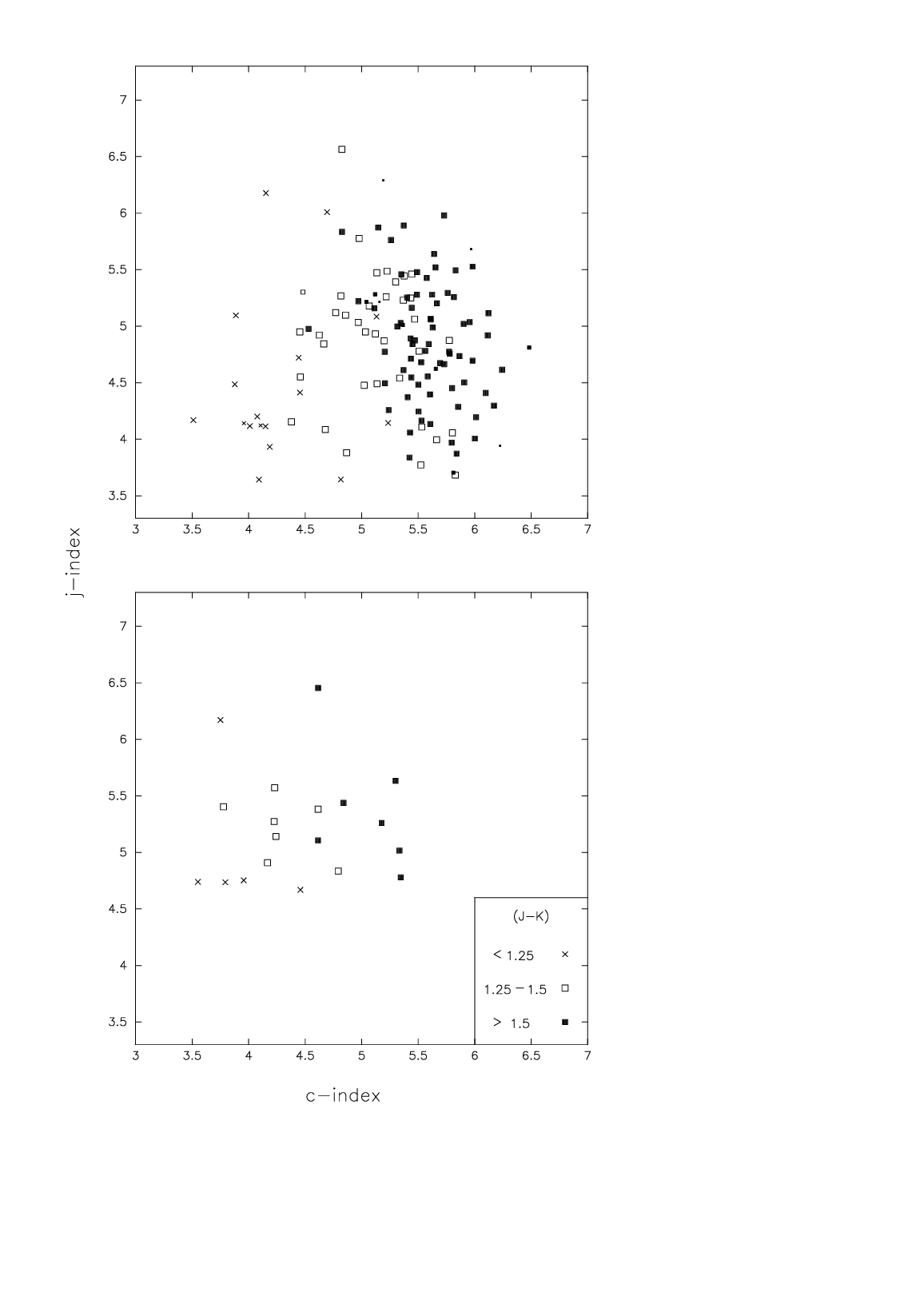}
\caption{The c-index plotted against j-index.  The upper panel is for 
the normal J~stars and the lower panel is for the bright J~stars.  
In each case the symbols mark ranges in $(J-K)$ as shown in the inset.}
\label{cind_jind}
\end{figure}

Fig.~\ref{cind_jind} shows the distribution of J~stars according to 
c-index and j-index.   The upper panel is for the normal J~stars and 
the lower panel is for the bright J~stars.  In each case the symbols
mark ranges in $(J-K)$.  For the normal (fainter) J~stars, the reddest stars
represented by the filled squares all have strong carbon bands and cover the
full range of j-index.  The bluer stars (open square then cross) have 
progressively weaker carbon bands and fewer representatives with a 
high j-index.  The bright J~stars generally have higher j-indices and 
weaker c-indices than normal J~stars with the same $(J-K)$ colour.

\subsection{Statistics}

The number of stars observed with 2dF (excluding repeats) that gave spectra
good enough to be given the first-pass classification was 1497.  156 of 
these are J~stars which is ten per cent of the total.   This is somewhat 
less than the fraction of $\sim$15 per cent \cite{abia00} seen in 
the Galaxy.  However, this difference cannot be deemed to be significant 
until a full comparison of the selection effects is made.  The set of 
J~stars are divided into their six spectral groups in the following 
percentages: 12, 26, 2, 38, 11 and 11. The bright J~stars
make up 13 per cent of the total.  Just over two thirds of the bright
J~stars belong to Group~1 and vice versa.

\subsection{Location within LMC}

\begin{figure}
\vspace{15.0cm}
\includegraphics{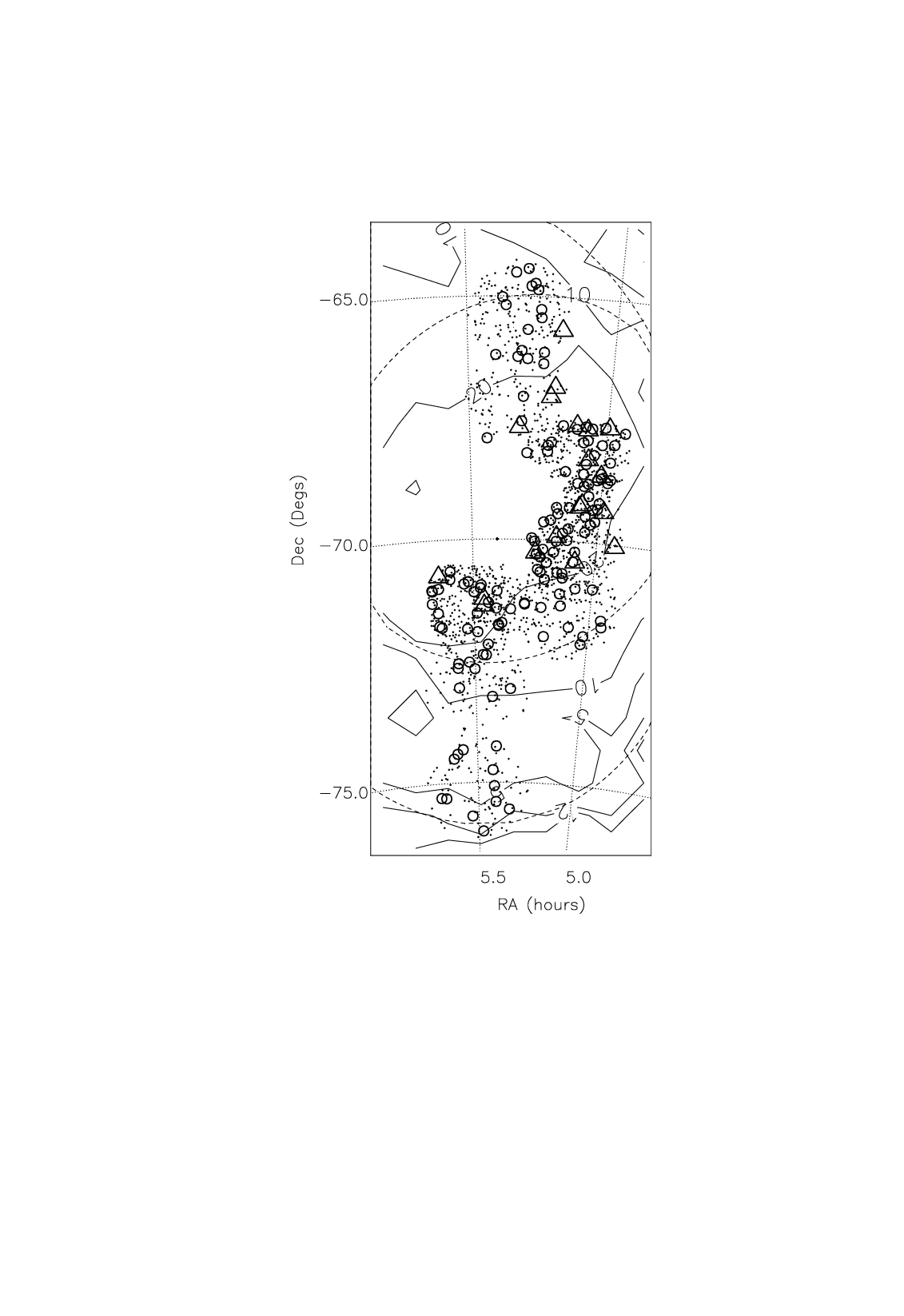}
\caption{All the carbon stars observed with 2dF plotted as a function of 
position.  The circles are the J~stars of spectral Groups 2-6 and the 
triangles are the J~stars of spectral Group~1.  Superimposed are labelled
carbon star surface density contours and two ellipse portions (dashed lines)
taken from the cluster distribution (see text).  The surface density units 
are the numbers of carbon stars per $40\times40$\,arcmin$^2$ box. }
\label{ra_dec}
\end{figure}

Fig.~\ref{ra_dec} shows all the carbon stars observed with 2dF plotted
according to their position in the LMC, with the J~stars identified and 
the Group~1 stars specially marked.  Superimposed are the numbered 
surface density contours of the carbon stars in KDMK01.  The surface 
density units are the numbers of carbon stars per $40\times40$\,arcmin$^2$ 
box.  Note that the catalogue is known to be incomplete in the very crowded 
region of the Bar.  The cluster systems of the LMC
appear to be found in a low density elliptical system offset to the east
of the centre of the Bar and a more centrally located higher density
region \cite{kontizas90}; so too are the planetary nebulae \cite{morgan94}.
Fig.~\ref{ra_dec} also shows two ellipses identifying these zones as 
determined from the clusters.  It is within the inner system that all 
the young, metal-rich clusters are found 
(Kontizas, Kontizas \& Michalitsianos \shortcite{kontizas93}, 
Kontizas et al. \shortcite{kontizas01a}).  It is clear that Group~1 stars are 
concentrated on the Bar and are absent from the southern part of the 
Cloud, and possibly from the NE part as well.  No other spectral group
shows this kind of behaviour.  All but two of the nineteen Group~1 J~stars 
are found in the central part of the LMC as defined by the contour 
labelled `20' and comprise 17 per cent of the total number of J~stars 
found there.  If the population mix were uniform throughout the LMC,
then the probability of obtaining this result is $<$0.01 whether considering
the detection of no Group~1 stars beyond the `10'-contour or two
beyond the `20'-contour.   The location of one Group~1 star 
in the north suggests that Group~1 stars might be associated with the 
inner elliptical system rather than the Bar itself.  

Similar exercises can be carried out with the other parameters. It turns 
out that none of the bright J~stars are located in the south of the LMC
(Dec $ < -72^\circ$).  The c-index, when plotted against Dec, also shows 
some variation with position in that the stars with a low c-index are 
also concentrated towards the central regions.  These two considerations 
produce essentially the same result as noted in the paragraph above because 
70 per cent of the bright stars are Group~1 stars and the bright stars 
generally have low levels of the c-index.   No conclusions can be drawn for
the bright stars or those with low c-index if the Group~1 stars are omitted
because there are too few.   There is no such result for 
the j-indices because the most common stars - the red end of the
J-star sequence - are spread across all declinations and exhibit
the full range of j-index (see Fig.~\ref{cind_jind}).

\subsection{CH stars}

Hartwick \& Cowley \shortcite{hartwick88} and Cowley \& Hartwick
\shortcite{cowley91} have published lists of candidate CH~stars in the LMC.
It has been noted that most of these stars lie to 
the bright and blue side of the N~stars in the $(J-K) - K$ colour-magnitude 
diagram \cite{feast92}, a point that has led Suntzeff et al. 
\shortcite{suntzeff93} to conclude that they are not members of the 
old stellar population in the LMC analagous to the Galactic Population II 
CH~stars, but relatively young carbon stars.   Since the Group~1 J~stars 
have similar photometric characteristics, it is 
worth considering whether the Group~1 J~stars and the Hartwick-Cowley
`CH~stars' are the same.  Although most of the Hartwick-Cowley stars 
lie in the outer parts of the LMC beyond the limits of the 2dF observations,
eight were observed with 2dF: six are J~stars 
(HC\,143, HC\,153, HC\,160, HC\,185, HC\,192, 
and HC\,198) and two are not (HC\,200, HC\,206).  Two of these J~stars 
belong to Group~1. Also, one (HC\,165)
has been reported to be a J~star by Richer et al. \shortcite{richer79}.    
With seven out of nine being J~stars, the Hartwick-Cowley 
stars could easily be a sample of J~stars, but not specifically a sample 
of Group~1 stars.  Although isotope ratios 
in CH~stars cover a wide range \cite{vanture92} overlapping that of 
the J~stars, confusion between the two classes exists only at the 
`marginal J' level \cite{barnbaum96}.  The Hartwick-Cowley stars 
seen by 2dF to be J~stars are by no means `marginal'.  The J~stars do have 
a depression between 4330 and 4380 but the shape is quite different 
from the P-branch at 4352\AA \/ which is characteristic of CH~stars
(see Barnbaum et al. \shortcite{barnbaum96}).  HC\,185 (KDM\,1814)
is the one star from the Hartwick-Cowley sample that was observed in 
the blue.  Its spectrum between 4250\AA \/ and 4450\AA \/ matches well 
with other J~stars such as NQ~Cas, given that its carbon bands are 
relatively weak (c-index\,=\,3.5); but there is no evidence of the
4352\AA \/ CH P-branch.

Six of the Group~1 stars were included in the sample of stars observed 
in the blue.  These spectra were then examined to see whether they showed 
evidence of strong CH-bands.   One spectrum is too weak to provide a 
good signal, but four of the others (KDM\,2312, 
KDM\,1814, KDM\,(1802 and KDM\,1249) 
have strong G-bands and the spectral depression between 
4350\AA \/ and 4380\AA \/ which is typical of J~stars, but without 
the bandhead of the P-branch at 4352\AA.  The 
sixth Group~1 star (KDM\,2832) has a weak G-band and no sign 
of the P-band.  
Nor did any show exhanced Ba\,{\sc ii}\,$\lambda$4554 or 
Ba\,{\sc ii}\,$\lambda$4934 as is usually seen in CH~stars.  
These considerations support the conclusion that the Group~1 J~stars 
are not CH~stars.

\section{Summary} 

\noindent (1) \hspace{0.2cm} A large sample of 1497 LMC carbon stars 
has been observed using the 
2dF facility on the AAT.  Of these, 156 (10 per cent) have been 
identified as J~stars.  This is less than the fraction of
$\sim$15 per cent \cite{abia00} seen in the Galaxy, but selection effects
could be important.

\noindent (2) \hspace{0.2cm} The spectra of the J~stars are varied in 
appearance and have been subdivided
into six spectral groups according to the 
slopes between the maxima at 6134\AA, 6180\AA \/ and 6202\AA.

\noindent (3) \hspace{0.2cm} In the $K - (J-K)$ colour magnitude diagram, 
most of the J~stars lie 
along a sequence parallel to and fainter than the sequence of N~stars.
Some stars (13 per cent) are brighter than this sequence: these are 
mostly of spectral Group~1.  This is contary to the older notion based
on shallower samples that most of the J~stars were brighter than the 
N~stars. 

\noindent (4) \hspace{0.2cm} A c-index and a j-index was assigned to 
each star to measure the strength of the C$_2$ bands and the relative 
strengths of the isotopic bands.
The spectral groupings are related to these indices.  For the 
common stars on the J-star photometric sequence, the redder ones have 
high c-indices and cover the full range of j-index.
The bluer stars have lower c-indices and fewer representatives 
with high j-indices.  The bright stars generally have higher j-indices
and lower c-indices than their fainter couterparts of the same
$(J-K)$ colour.

\noindent (5) \hspace{0.2cm} The $^{13}{\rmn C}^{13}{\rmn C}~\lambda6144$ 
band is strong in 
most stars with a high j-index, but is not always seen as an 
unblended feature.

\noindent (6) \hspace{0.2cm} The ratio of the 
$^{12}{\rmn C}^{14}{\rmn N}~\lambda6202$ and
$^{12}{\rmn C}^{12}{\rmn C}~\lambda6192$ bands is higher for the bright 
J~stars than for those on the principal J-star sequence.

\noindent (7) \hspace{0.22cm} The Na~D lines are weaker in LMC J~stars 
than in Galactic J~stars, and, as in the Galaxy, are weaker in the J~stars 
than in the N~stars.  The weakness of the Na~D lines in the LMC is due 
to the lower metallicity there.  The strengths of the Na~D lines do not
correlate with temperature as measured by ($J-K$).

\noindent (8) \hspace{0.2cm} Group~1 J~stars are absent from the southern 
parts of the LMC and found in the inner regions where the young, metal-rich 
clusters are found.  Bright J~stars and those with low c-index are also
concentrated towards the centre of the LMC, but no such effect is seen 
for the j-index.

\noindent (9) \hspace{0.2cm} The small overlap between the 2dF data and
the suspected CH~stars of Hartwick and Cowley suggests that the latter 
include a high percentage of J~stars.

\section*{Acknowledgements}

The authors are grateful to the staff of the Anglo-Australian Observatory
for assistance with the observations, to the Australian Time Assignment
Committee for the allocation of telescope time, and to the 2-MASS, DENIS, 
CDS and IPAC project teams for making the infrared photometry available via 
the WWW.

\label{lastpage}

 \begin{table*}
 \begin{center}
 \caption{J stars in the Large Magellanic Cloud}
 \label{data}
 \begin{tabular}{rp{0.2cm}cccccccccc}
 \hline
 KDM & & R & I & J & H & K & Group & c-index & j-index & 
$^{13}{\rmn C}^{13}{\rmn C}$ & Note \\
 & & & & & & & & & & &\\
 985 &&  14.64 &  13.91 &  12.61 &  11.87 &  11.54 &   1 &   3.8 &   4.7 &   0 &             \\
1008 &&  15.27 &  14.17 &  12.66 &   0.00 &  11.14 &   4 &   5.3 &   5.0 &   0 &             \\
1048 &&  15.02 &  14.02 &  12.49 &  11.38 &  10.70 &   4 &   5.5 &   4.5 &   0 &             \\
1067 &&  15.54 &  14.45 &  13.06 &  11.86 &  11.29 &   5 &   5.8 &   5.5 &   1 &             \\
1156 &&  15.36 &  14.02 &  13.58 &  12.81 &  12.46 &   2 &   4.2 &   3.9 &   0 &             \\
1169 &&  14.69 &  13.68 &  12.28 &  11.18 &  10.70 &   2 &   4.8 &   5.8 &   0 &             \\
1215 &&  14.78 &  13.96 &  12.97 &  12.01 &  11.68 &   2 &   5.2 &   5.3 &   0 &             \\
1232 &&  15.44 &  14.36 &  12.76 &  11.40 &  10.80 &   5 &   5.2 &   6.3 &   0 & $\ddagger$  \\
1249 &&  14.38 &  13.58 &  11.97 &  11.03 &  10.56 &   1 &   3.8 &   5.4 &   1 &             \\
1290 &&  14.95 &  14.06 &  12.97 &  12.07 &  11.80 &   1 &   3.9 &   4.5 &   0 &             \\
1327 &&  14.70 &  13.79 &  12.16 &  10.99 &  10.51 &   2 &   5.4 &   4.9 &   0 &             \\
1365 &&  15.19 &  14.28 &  13.28 &  12.21 &  11.77 &   2 &   5.1 &   5.2 &   0 &             \\
1372 &&  15.14 &  13.99 &  12.84 &  11.77 &  11.22 &   4 &   5.0 &   5.2 &   0 & $\dagger$   \\
1378 &&  15.32 &  14.07 &  12.97 &  12.01 &  11.61 &   1 &   4.8 &   5.1 &   0 &             \\
1380 &&  16.13 &  15.19 &  14.21 &  13.47 &  13.17 &   2 &   4.1 &   4.2 &   0 &             \\
1389 &&  14.87 &  13.88 &  12.51 &  11.43 &  10.86 &   4 &   5.7 &   4.7 &   0 &             \\
1393 &&  14.89 &  13.86 &  12.58 &  11.31 &  10.74 &   5 &   5.4 &   5.9 &   2 &             \\
1410 &&  15.28 &  13.89 &  12.58 &  11.49 &  10.96 &   4 &   5.8 &   3.9 &   0 &             \\
1417 &&  14.35 &  13.31 &  11.78 &  10.59 &  10.09 &   2 &   5.2 &   5.3 &   1 &             \\
1450 &&  14.76 &  13.89 &  12.83 &  11.56 &  11.02 &   4 &   5.4 &   5.0 &   0 & $\dagger$   \\
1456 &&  15.37 &  14.31 &  12.75 &  11.58 &  10.99 &   4 &   5.8 &   4.3 &   0 &             \\
1504 &&  14.85 &  14.01 &  13.00 &  12.04 &  11.73 &   2 &   5.1 &   5.2 &   0 &             \\
1510 &&  15.37 &  14.19 &  12.76 &  11.67 &  11.25 &   4 &   5.6 &   4.8 &   0 &             \\
1529 &&  15.78 &  14.53 &  13.12 &  11.94 &  11.33 &   4 &   5.4 &   4.6 &   0 &             \\
1567 &&  15.11 &  14.16 &  13.41 &  12.34 &  11.89 &   4 &   5.5 &   4.2 &   0 &             \\
1608 &&  15.02 &  14.30 &  13.54 &  12.73 &  12.46 &   4 &   4.8 &   3.6 &   0 &             \\
1637 &&  15.55 &  14.79 &  13.87 &  13.02 &  12.73 &   2 &   4.0 &   4.1 &   0 &             \\
1641 &&  14.77 &  12.96 &  12.96 &  11.94 &  11.55 &   2 &   5.1 &   4.5 &   0 &             \\
1653 &&  15.65 &  14.76 &  13.64 &  12.90 &  12.60 &   2 &   4.2 &   4.1 &   0 &             \\
1660 &&  15.44 &  14.44 &  12.95 &  11.79 &  11.27 &   2 &   5.2 &   4.8 &   0 &             \\
1688 &&  14.26 &  13.46 &  11.92 &  10.73 &  10.24 &   1 &   4.8 &   5.4 &   0 & *           \\
1736 &&  15.21 &  14.21 &  13.07 &  12.04 &  11.62 &   4 &   5.8 &   3.7 &   0 &             \\
1737 &&  15.14 &  14.14 &  12.63 &  11.59 &  11.20 &   4 &   5.8 &   4.1 &   0 &             \\
1742 &&  14.87 &  14.15 &  13.51 &  12.65 &  12.33 &   2 &   4.7 &   6.0 &   0 &             \\
1745 &&  14.60 &  13.94 &  13.01 &  12.19 &  11.96 &   1 &   3.8 &   6.2 &   1 &             \\
1751 &&  13.72 &  12.47 &  11.31 &  10.27 &   9.73 &   1 &   4.6 &   5.1 &   0 & *           \\
1762 &&  14.98 &  14.13 &  12.66 &  11.66 &  11.21 &   4 &   5.4 &   5.4 &   0 &             \\
1767 &&  15.08 &  14.02 &  12.73 &  11.53 &  10.93 &   4 &   5.8 &   4.0 &   0 &             \\
1796 &&  15.66 &  14.38 &  12.73 &  11.48 &  10.82 &   4 &   5.9 &   5.0 &   0 &             \\
1802 &&  13.95 &  13.36 &  11.94 &  10.93 &  10.62 &   1 &   4.2 &   5.3 &   1 &             \\
1814 &&  13.61 &  13.05 &  11.26 &  10.47 &  10.18 &   1 &   3.5 &   4.7 &   0 & *           \\
1820 &&  15.54 &  14.41 &  13.03 &  11.82 &  11.24 &   6 &   6.1 &   5.1 &   0 &             \\
1849 &&  14.86 &  13.99 &  12.56 &  11.55 &  11.03 &   4 &   5.2 &   4.5 &   0 &             \\
1864 &&  15.08 &  14.27 &  13.21 &  12.40 &  12.14 &   2 &   4.5 &   4.7 &   0 &             \\
1880 &&  15.08 &  14.35 &  12.82 &  11.86 &  11.44 &   4 &   5.4 &   5.2 &   1 &             \\
1901 &&  15.66 &  14.75 &  13.46 &  12.56 &  12.19 &   4 &   4.5 &   4.6 &   0 &             \\
1988 &&  15.31 &  14.12 &  12.71 &  11.52 &  10.99 &   4 &   5.8 &   5.3 &   0 &             \\
1989 &&  15.59 &  14.86 &  14.30 &  13.51 &  13.32 &   1 &   4.1 &   4.1 &   0 & $\dagger$   \\
2035 &&  14.82 &  13.91 &  13.01 &  11.97 &  11.64 &   2 &   5.5 &   5.1 &   0 &             \\
2039 &&  14.85 &  13.89 &  13.16 &  12.04 &  11.57 &   4 &   5.6 &   4.4 &   0 &             \\
2043 &&  14.33 &  13.52 &  12.79 &  11.76 &  11.37 &   2 &   5.3 &   5.4 &   0 &             \\
2098 &&  15.46 &  14.27 &  13.07 &  11.80 &  11.15 &   6 &   5.8 &   4.8 &   0 &             \\
2099 &&  15.33 &  14.25 &  12.68 &  11.57 &  10.99 &   4 &   5.4 &   4.7 &   0 &             \\
2112 &&  15.02 &  14.24 &  13.22 &  12.38 &  11.95 &   2 &   4.7 &   4.8 &   0 &             \\
2135 &&  15.76 &  14.56 &  13.31 &  12.09 &  11.45 &   4 &   5.6 &   5.1 &   0 &             \\
2155 &&  15.19 &  14.18 &  12.61 &  11.44 &  10.83 &   5 &   5.7 &   6.0 &   2 &             \\
 \hline
 \end{tabular}
 \end{center}
 \end{table*}
 \begin{table*}
 \begin{center}
 \contcaption{J stars in the Large Magellanic Cloud}
 \begin{tabular}{rp{0.2cm}cccccccccc}
 \hline
 KDM & & R & I & J & H & K & Group & c-index & j-index & 
$^{13}{\rmn C}^{13}{\rmn C}$ & Note \\
 & & & & & & & & & & &\\
2184 &&  14.74 &  13.80 &  12.50 &  11.31 &  10.71 &   6 &   6.5 &   4.8 &   0 & $\dagger$   \\
2244 &&  15.18 &  14.17 &  13.15 &  11.95 &  11.41 &   6 &   5.4 &   5.2 &   1 &             \\
2293 &&  15.88 &  14.63 &  14.20 &  13.30 &  13.07 &   2 &   4.4 &   4.4 &   0 &             \\
2310 &&  14.70 &  13.89 &  13.16 &   0.00 &  11.49 &   4 &   5.4 &   4.6 &   0 &             \\
2312 &&  14.69 &  13.97 &  13.17 &  12.31 &  12.07 &   1 &   4.2 &   6.2 &   0 &             \\
2313 &&  14.74 &  13.78 &  12.46 &  11.33 &  10.89 &   4 &   6.1 &   4.4 &   0 &             \\
2348 &&  15.07 &  13.89 &  12.61 &   0.00 &  11.00 &   5 &   5.7 &   5.5 &   2 &             \\
2364 &&  15.44 &  14.38 &  13.04 &  11.85 &  11.20 &   5 &   5.5 &   5.3 &   1 &             \\
2419 &&  14.79 &  13.74 &  11.74 &   0.00 &  10.21 &   1 &   4.6 &   6.4 &   2 & *           \\
2475 &&  14.98 &   0.00 &  12.70 &  11.57 &  10.96 &   4 &   5.9 &   4.5 &   0 &             \\
2517 &&  15.48 &  14.18 &  12.87 &  11.66 &  11.07 &   6 &   6.0 &   4.2 &   0 &             \\
2518 &&  13.33 &  12.63 &  10.93 &  10.02 &   9.60 &   1 &   4.2 &   4.9 &   0 & *           \\
2543 &&  14.95 &  14.17 &  12.85 &  11.86 &  11.45 &   4 &   5.3 &   4.5 &   0 &             \\
2561 &&  15.14 &  14.20 &  12.88 &  11.89 &  11.60 &   2 &   4.6 &   4.9 &   0 &             \\
2573 &&  14.84 &  13.92 &  12.64 &  11.65 &  11.36 &   2 &   4.9 &   5.1 &   0 & *           \\
2614 &&  15.06 &  14.04 &  12.32 &  11.28 &  10.88 &   1 &   4.5 &   5.3 &   0 & $\dagger$   \\
2622 &&  15.03 &  14.06 &  12.56 &  11.45 &  10.86 &   5 &   5.2 &   5.9 &   1 &             \\
2630 &&  15.04 &  14.18 &  12.79 &  11.74 &  11.39 &   2 &   5.0 &   4.9 &   0 &             \\
2635 &&  15.32 &  14.31 &  12.69 &  11.62 &  11.08 &   4 &   5.3 &   5.0 &   0 &             \\
2646 &&  15.65 &  13.85 &  13.06 &  11.91 &  11.37 &   4 &   5.8 &   4.8 &   0 &             \\
2654 &&  15.02 &  13.88 &  12.88 &  11.91 &  11.39 &   4 &   5.7 &   4.0 &   0 &             \\
2696 &&  14.68 &  13.69 &  12.29 &  11.17 &  10.73 &   5 &   5.6 &   5.1 &   1 &             \\
2708 &&  14.72 &  13.75 &  12.48 &  11.26 &  10.66 &   4 &   4.5 &   5.0 &   0 &             \\
2752 &&  15.89 &  14.79 &  13.44 &   0.00 &  11.83 &   6 &   6.2 &   4.6 &   0 &             \\
2803 &&  15.48 &  14.15 &  12.70 &   0.00 &   0.17 &   6 &   6.1 &   4.4 &   0 & $\dagger$   \\
2806 &&  14.78 &  13.66 &  11.72 &  10.53 &  10.06 &   5 &   0.0 &   0.0 &   0 & $\ddagger$  \\
2807 &&  15.11 &  14.10 &  12.66 &  11.51 &  10.96 &   4 &   5.4 &   5.2 &   0 &             \\
2832 &&  14.60 &  13.71 &  11.76 &  10.74 &  10.34 &   1 &   4.2 &   5.6 &   1 &             \\
2868 &&  15.86 &  14.92 &  14.36 &  13.63 &  13.44 &   2 &   3.5 &   4.2 &   0 &             \\
2882 &&  15.26 &  14.08 &  12.78 &  11.60 &  10.92 &   4 &   5.4 &   4.8 &   0 &             \\
2897 &&  15.57 &  14.10 &  12.18 &  11.09 &  10.59 &   5 &   5.6 &   5.6 &   1 &             \\
2946 &&  15.19 &  14.02 &  12.87 &  11.75 &  11.27 &   3 &   5.2 &   4.3 &   0 &             \\
2969 &&  15.35 &  14.17 &  12.56 &  11.32 &  10.64 &   5 &   5.8 &   5.3 &   1 &             \\
3036 &&  15.25 &  14.12 &  12.92 &  11.69 &  11.13 &   4 &   5.5 &   4.9 &   0 & $\dagger$   \\
3090 &&  14.87 &  14.06 &  12.47 &   0.00 &  10.92 &   5 &   5.7 &   5.2 &   2 &             \\
3093 &&  14.97 &  14.25 &  13.42 &   0.00 &  12.39 &   2 &   3.9 &   5.1 &   0 &             \\
3124 &&  14.63 &  13.55 &  11.72 &  10.67 &  10.10 &   5 &   5.3 &   5.6 &   2 & *           \\
3135 &&  15.35 &  14.18 &  12.47 &   0.00 &  10.66 &   4 &   5.2 &   5.2 &   0 & $\ddagger$  \\
3161 &&  15.32 &  14.30 &  12.67 &   0.00 &  11.10 &   2 &   5.1 &   5.3 &   2 & $\dagger$   \\
3171 &&  15.65 &  14.28 &  12.94 &  11.57 &  10.83 &   6 &   5.8 &   4.4 &   0 & $\dagger$   \\
3178 &&  15.04 &  14.03 &  12.53 &  11.59 &  11.25 &   2 &   4.8 &   4.8 &   0 & *           \\
3226 &&  15.18 &  14.10 &  12.48 &   0.00 &  11.03 &   2 &   5.1 &   5.5 &   0 &             \\
3244 &&  14.97 &  13.78 &  12.82 &   0.00 &  11.44 &   4 &   5.4 &   5.2 &   1 &             \\
3263 &&  15.32 &  14.49 &  13.70 &   0.00 &  12.50 &   4 &   5.2 &   4.1 &   0 &             \\
3283 &&  14.94 &  14.02 &  12.84 &   0.00 &  11.52 &   1 &   4.4 &   4.9 &   0 & $\dagger$   \\
3339 &&  15.19 &  14.17 &  13.00 &   0.00 &  11.54 &   2 &   5.1 &   4.9 &   0 &             \\
3362 &&  14.92 &  13.89 &  12.54 &  11.40 &  10.81 &   4 &   5.8 &   4.6 &   0 &             \\
3389 &&  15.17 &  14.20 &  12.67 &  11.69 &  11.31 &   3 &   4.7 &   4.1 &   0 &             \\
3391 &&  14.77 &  13.81 &  12.48 &  11.43 &  11.03 &   3 &   4.9 &   3.9 &   0 &             \\
3425 &&  14.59 &  13.72 &  12.46 &   0.00 &  11.07 &   2 &   5.0 &   5.8 &   0 & *           \\
3583 &&  15.64 &  14.56 &  13.73 &  12.68 &  12.19 &   4 &   5.8 &   3.7 &   0 & $\dagger$   \\
3643 &&  14.13 &  13.12 &  11.69 &  10.67 &  10.16 &   2 &   5.3 &   5.0 &   0 & *           \\
3655 &&  15.17 &  14.09 &  12.49 &  11.48 &  11.03 &   4 &   5.8 &   4.9 &   2 &             \\
3726 &&  14.56 &  13.74 &  12.45 &  11.47 &  11.06 &   2 &   4.8 &   6.6 &   1 & *           \\
3735 &&  15.19 &  14.33 &  13.25 &  12.33 &  11.90 &   2 &   5.0 &   4.5 &   0 &             \\
3762 &&  14.84 &  13.92 &  12.73 &  11.48 &  10.89 &   5 &   6.0 &   5.7 &   1 & $\ddagger$  \\
 \hline
 \end{tabular}
 \end{center}
 \end{table*}
 \begin{table*}
 \begin{center}
 \contcaption{J stars in the Large Magellanic Cloud}
 \begin{tabular}{rp{0.2cm}cccccccccc}
 \hline
 KDM & & R & I & J & H & K & Group & c-index & j-index & 
$^{13}{\rmn C}^{13}{\rmn C}$ & Note \\
 & & & & & & & & & & &\\
3786 &&  14.86 &  14.05 &  12.24 &  11.26 &  10.76 &   2 &   4.4 &   4.2 &   0 &             \\
3787 &&  15.42 &  14.24 &  12.72 &   0.00 &  11.06 &   4 &   5.7 &   4.6 &   0 & $\dagger$   \\
3794 &&  15.30 &  14.20 &  12.50 &   0.00 &  10.59 &   5 &   5.5 &   5.5 &   1 &             \\
3809 &&  15.24 &  14.36 &  13.32 &  12.37 &  11.97 &   4 &   5.2 &   4.9 &   0 & *           \\
3865 &&  15.19 &  14.17 &  13.03 &  11.81 &  11.16 &   6 &   5.5 &   4.7 &   0 &             \\
3908 &&  15.40 &  14.38 &  13.35 &  12.38 &  11.97 &   4 &   5.5 &   4.1 &   0 &             \\
3909 &&  15.76 &  14.46 &  13.03 &  11.84 &  11.24 &   6 &   5.6 &   5.0 &   1 &             \\
3995 &&  15.42 &  14.48 &  12.89 &  11.79 &  11.34 &   4 &   6.2 &   3.9 &   0 & $\ddagger$  \\
3998 &&  15.29 &  13.93 &  12.69 &  11.42 &  10.65 &   6 &   5.6 &   4.1 &   0 &             \\
4022 &&  15.75 &  14.60 &  13.23 &  12.22 &  11.78 &   4 &   5.5 &   4.8 &   0 &             \\
4081 &&  15.44 &  14.53 &  13.13 &  12.24 &  11.90 &   2 &   4.4 &   4.7 &   0 &             \\
4129 &&  14.56 &  13.75 &  12.69 &  11.83 &  11.54 &   1 &   4.0 &   4.8 &   0 &             \\
4141 &&  14.40 &  13.60 &  12.06 &  10.93 &  10.56 &   1 &   4.6 &   5.4 &   1 &             \\
4189 &&  15.40 &  14.34 &  13.46 &  12.60 &  12.16 &   4 &   5.5 &   3.8 &   0 &             \\
4219 &&  14.52 &  13.65 &  11.95 &  10.89 &  10.40 &   2 &   5.3 &   4.8 &   0 & *           \\
4236 &&  15.42 &  14.23 &  12.61 &  11.40 &  10.81 &   5 &   5.3 &   5.5 &   1 &             \\
4271 &&  15.28 &  14.47 &  13.19 &  12.20 &  11.77 &   2 &   5.4 &   5.5 &   0 &             \\
4328 &&  14.89 &  13.85 &  12.65 &  11.47 &  10.94 &   4 &   5.0 &   5.2 &   1 &             \\
4332 &&  15.24 &  14.38 &  13.03 &  11.89 &  11.50 &   4 &   5.6 &   4.6 &   0 &             \\
4433 &&  14.47 &  14.46 &  13.10 &  11.88 &  11.36 &   6 &   5.7 &   4.7 &   0 &             \\
4453 &&  15.10 &  14.05 &  12.68 &  11.62 &  11.08 &   4 &   5.5 &   4.2 &   0 &             \\
4582 &&  14.86 &  13.67 &  12.53 &  11.22 &  10.67 &   6 &   6.0 &   4.7 &   0 &             \\
4626 &&  15.91 &  14.34 &  12.82 &  11.50 &  10.84 &   6 &   5.6 &   4.8 &   1 &             \\
4653 &&  15.11 &  13.98 &  12.60 &  11.38 &  10.85 &   5 &   5.3 &   5.8 &   2 &             \\
4675 &&  15.01 &  14.02 &  12.45 &  11.32 &  10.75 &   5 &   5.4 &   5.8 &   1 &             \\
4708 &&  15.91 &  14.34 &  13.24 &  11.98 &  11.35 &   4 &   6.2 &   4.3 &   0 &             \\
4944 &&  14.89 &  14.03 &  13.19 &  12.20 &  11.75 &   2 &   5.2 &   5.5 &   1 &             \\
4956 &&  15.21 &  14.15 &  12.99 &  11.81 &  11.31 &   6 &   6.0 &   5.5 &   1 &             \\
4976 &&  15.24 &  14.41 &  13.59 &  12.79 &  12.46 &   2 &   4.1 &   3.6 &   0 &             \\
5094 &&  15.08 &  14.26 &  13.34 &  12.46 &  12.11 &   4 &   5.1 &   5.1 &   1 &             \\
5111 &&  14.91 &  13.94 &  12.73 &  11.57 &  11.03 &   4 &   6.0 &   4.0 &   0 &             \\
5122 &&  15.14 &  14.23 &  13.02 &  11.96 &  11.56 &   4 &   0.0 &   0.0 &   0 & $\ddagger$  \\
5246 &&  15.01 &  13.90 &  12.19 &  11.09 &  10.47 &   4 &   5.6 &   5.3 &   1 &             \\
5368 &&  14.87 &  13.78 &  12.21 &  11.07 &  10.52 &   4 &   5.4 &   4.4 &   0 &             \\
5391 &&  13.95 &  13.00 &  11.80 &  10.79 &  10.43 &   1 &   4.2 &   5.1 &   1 &             \\
5405 &&  15.54 &  14.11 &  12.25 &  10.93 &  10.22 &   6 &   6.1 &   4.9 &   1 &             \\
5433 &&  14.99 &  14.10 &  12.94 &  11.98 &  11.62 &   2 &   5.0 &   5.0 &   0 &             \\
5446 &&  15.69 &  14.55 &  13.32 &  12.22 &  11.65 &   4 &   5.9 &   4.7 &   0 &             \\
5544 &&  15.34 &  14.34 &  13.31 &  12.15 &  11.59 &   6 &   6.0 &   5.0 &   0 &             \\
5562 &&  14.94 &  13.88 &  12.83 &  11.89 &  11.50 &   2 &   4.8 &   5.3 &   0 &             \\
5574 &&  15.02 &  14.02 &  12.83 &  11.75 &  11.24 &   4 &   5.4 &   4.1 &   0 &             \\
5585 &&  15.23 &  14.20 &  12.64 &  11.58 &  11.03 &   4 &   5.4 &   3.8 &   0 &             \\
5703 &&  14.97 &   0.00 &  12.79 &  11.78 &  11.28 &   4 &   5.6 &   5.4 &   0 &             \\
6656 &&  15.33 &  14.48 &  13.63 &  12.79 &  12.41 &   2 &   4.0 &   4.1 &   0 & $\dagger$   \\
 \hline
 \multicolumn{11}{l}{Notes: $\dagger$ and
  $\ddagger$ -- weaker spectra (see text);
 * -- a cross-identification noted below} \\
 \multicolumn{11}{l}{KDM\,1688 $\equiv$ WORC\,70} \\
 \multicolumn{11}{l}{KDM\,1751 $\equiv$ WORC\,74} \\
 \multicolumn{11}{l}{KDM\,1814 $\equiv$ 
 WORC\,78 $\equiv$ CRW\,6-6 $\equiv$ HC\,185} \\
 \multicolumn{11}{l}{KDM\,2419 $\equiv$ WORC\,101} \\
 \multicolumn{11}{l}{KDM\,2518 $\equiv$ WORC\,106 $\equiv$ HC\,160} \\
 \multicolumn{11}{l}{KDM\,2573 $\equiv$ WORC\,109} \\
 \multicolumn{11}{l}{KDM\,3124 $\equiv$ WORC\,131 $\equiv$ CRW\,4-9} \\
 \multicolumn{11}{l}{KDM\,3178 $\equiv$ HC\,153} \\
 \multicolumn{11}{l}{KDM\,3425 $\equiv$ HC\,143} \\
 \multicolumn{11}{l}{KDM\,3643 $\equiv$ WORC\,146} \\
 \multicolumn{11}{l}{KDM\,3726 $\equiv$ HC\,198} \\
 \multicolumn{11}{l}{KDM\,3809 $\equiv$ HC\,192} \\
 \multicolumn{11}{l}{KDM\,4219 $\equiv$ WORC\,168} \\
 \end{tabular}
 \end{center}
 \end{table*}

\end{document}